 \documentclass[11pt]{article} 
 \usepackage{a4wide}  

\RequirePackage[latin1]{inputenc}     
 
\usepackage[english]{babel} 
\usepackage{amsmath} 
\usepackage{amsfonts} 
\usepackage{amssymb} 
\usepackage{xspace} 
\usepackage{latexsym} 
\usepackage{url} 
\usepackage{xspace} 
\usepackage{fancyvrb} 

\usepackage{graphics,color}              
\RequirePackage[latin1]{inputenc}   
 
\newenvironment{restate-proposition}[2][{}]{\noindent\textbf{Proposition~{#2}}\;\textbf{#1}\  
}{\vskip 1em} 
 
\newenvironment{restate-theorem}[2][{}]{\noindent\textbf{Theorem~{#2}}\;\textbf{#1}\  
}{\vskip 1em} 
 
\newenvironment{restate-corollary}[2][{}]{\noindent\textbf{Corollary~{#2}}\;\textbf{#1}\  
}{\vskip 1em}

\newcommand{\Proofitem}[1]{\medskip \noindent $#1\;$} 
\newcommand{\Proofitemf}[1]{\noindent $#1\;$} 
\newcommand{\Defitem}[1]{\smallskip \noindent $#1\;$} 
 
\newcommand{\Defitemf}[1]{\noindent $#1\;$}

\newcommand{\hbra}{\noindent\hbox to \textwidth{\leaders\hrule height1.8mm depth-1.5mm\hfill}} 
\newcommand{\hket}{\noindent\hbox to \textwidth{\leaders\hrule height0.3mm\hfill}} 
\newcommand{\ratio}{.3}

\newtheorem{theorem}{Theorem} 
\newtheorem{fact}[theorem]{Fact} 
\newtheorem{definition}[theorem]{Definition} 
\newtheorem{lemma}[theorem]{Lemma} 
 
\newtheorem{proposition}[theorem]{Proposition} 
\newtheorem{example}[theorem]{Example} 
 
\newtheorem{remark}[theorem]{Remark}

\newcommand{\qed}{\hfill${\Box}$}

\newcommand{\Figbar}{{\center \rule{\hsize}{0.3mm}}}    
 
\newcommand{\la}{\langle}               
\newcommand{\ra}{\rangle}

\newcommand{\ul}[1]{\underline{#1}}     
\newcommand{\ol}[1]{\overline{#1}}      
 
\newcommand{\Gives}{\vdash}             
 
\newcommand{\arrow}{\rightarrow}        
\newcommand{\Alt}{ \mid\!\!\mid  } 
\newcommand{\isum}{\oplus} 
 
\newcommand{\infer}[2]{\begin{array}{c} #1 \\ \hline #2 \end{array}} 
 
\newcommand{\Or}{\vee}                  
\newcommand{\AND}{\wedge}               
 
\newcommand{\wbis}{\approx}             
\newcommand{\twbis}{\approx^{t}}        

\newcommand{\union}{\cup}               
\newcommand{\inter}{\cap}               
\newcommand{\Union}{\bigcup}            
\newcommand{\minus}{\backslash}         
\newcommand{\comp}{\circ}               
\newcommand{\set}[1]{\{#1\}}            
 
\newcommand{\dcl}{\downarrow}           
\newcommand{\ucl}{\uparrow}             
\newcommand{\nor}{\succeq}
\newcommand{\rl}[1]{\;{\cal #1}\;}      
\newcommand{\rel}[1]{{\cal #1}}         
\newcommand{\trl}[2]{\;{\cal #1}_{#2}\;}
\newcommand{\trel}[2]{{\cal #1}_{#2}}   

\newcommand{\w}[1]{{\it #1}}    

\newcommand{\xst}[2]{\exists\, #1\;\: #2}

\newcommand{\s}[1]{{\sf #1}}    
\newcommand{\vc}[1]{{\bf #1}}

\newcommand{\act}[1]{\xrightarrow{#1}} 

\newcommand{\acteq}[1]{\stackrel{#1}{\leadsto}} 

\newcommand{\tact}[2]{\ensuremath{\underset{#1}{\act{~#2~}}}}
\newcommand{\tacteq}[2]{\ensuremath{\underset{#1}{\acteq{~#2~}}}}

\newcommand{\wact}[1]{\stackrel{#1}{\Rightarrow}} 

\newcommand{\twact}[2]{\ensuremath{\underset{#1}{\wact{~#2~}}}}

\newcommand{\eval}{\Downarrow}

\newcommand{\spi}{S\pi}

\newcommand{\emit}[2]{\ol{#1}#2}  
\newcommand{\present}[4]{#1(#2).#3,#4} 
\newcommand{\match}[4]{[#1=#2]#3,#4}       

\newcommand{\matchv}[4]{[#1 \unrhd #2]#3,#4}

\newcommand{\new}[2]{\nu #1 \ #2} 
\newcommand{\outact}[3]{\new{{\bf #1}}{\emit{#2}{#3}}} 
\newcommand{\real}{\makebox[5mm]{\,$\|\!-$}}

\newcommand{\om}[1]{\ensuremath{#1^\omega}}

\newcommand{\usage}[3]{\ensuremath{
    \begin{pmatrix}
      #1, #2, #3\\
    \end{pmatrix}
  }
}

\begin{document} 
 
\title{On affine usages in signal-based communication}

\author{Roberto M. Amadio
\quad Mehdi Dogguy \\
Universit\'e Paris Diderot (Paris 7), PPS, UMR-7126}

\maketitle 

\begin{abstract}
We describe a type system for a {\em synchronous} $\pi$-calculus
formalising the notion of {\em affine} usage in {\em signal-based}
communication.  In particular, we identify a limited number of usages
that preserve affinity and that can be composed. 
As a main application of the resulting system, we show
that typable programs are {\em deterministic}.
\end{abstract}

\section{Introduction}\label{intro-sec}
We are interested in {\em synchronous} systems.  In these systems, there is
a notion of {\em instant} (or phase, or pulse, or round) and at each
instant each component of the system, {\em a thread}, performs some
actions and synchronizes with all the other threads. One may say that all
threads proceed at the same speed and it is in this specific sense
that we shall refer to {\em synchrony} in this work.
{\em Signal-based} communication is often used as the basic interaction
mechanism in synchronous systems (see, {\em e.g.}, \cite{BG92,BD95}). 
Signals play a role similar to {\em channels} in asynchronous systems. 
Our goal in this paper is to study the notion of {\em affine usage} in
this context. In particular, we shall formalise our ideas in the
context of a {\em synchronous} $\pi$-calculus ($\spi$-calculus)
introduced in \cite{Amadio06}. We assume that the reader is familiar with the
$\pi$-calculus and proceed to give a flavour of the language
(the formal definition of the $\spi$-calculus is recalled 
in section \ref{section-spi}).

The syntax of the $\spi$-calculus is similar to the one of
the $\pi$-calculus, however there are some important {\em semantic}
differences that we highlight in the following simple example.
Assume $v_1\neq v_2$ are two distinct values and consider the
following program in $\spi$:
\[
P=\nu \ s_1,s_2 \ 
(\ \emit{s_{1}}{v_{1}}  \mid  
 \emit{s_{1}}{v_{2}}  \mid   
 s_1(x). \ (s_1(y). \  (s_2(z). \ A(x,y) \  \ul{,B(!s_1)}) 
 \ul{,0}) \ \ul{,0} \ )
\]
If we forget about the underlined parts and we regard $s_1,s_2$ as
{\em channel names} then $P$ could also be viewed as a $\pi$-calculus
process. In this case, $P$ would reduce to 
$
P_1 =  \new{s_1,s_2}{(s_2(z).A(\theta(x),\theta(y))}
$
where $\theta$ is a substitution such that
$\theta(x),\theta(y)\in \set{v_1,v_2}$ and $\theta(x)\neq \theta(y)$.
In $\spi$, {\em signals persist within the instant} and 
$P$ reduces to 
$
P_2 = \new{s_1,s_2}{(\emit{s_{1}}{v_{1}} \mid 
 \emit{s_{1}}{v_{2}} \mid (s_2(z).A(\theta(x),\theta(y))\ul{,B(!s_1)}))}
$
where  again $\theta(x),\theta(y)\in \set{v_1,v_2}$ but possibly
$\theta(x)=\theta(y)$.
What happens next? In the $\pi$-calculus, $P_1$ is 
{\em deadlocked} and no further computation is possible.
In the $\spi$-calculus, 
the fact that no further computation is possible in $P_2$ is
detected and marks the {\em end of the current instant}. Then
an additional computation represented by the relation $\act{N}$ 
moves $P_2$ to the following instant:
$
P_2 \act{N} P'_2 =  \new{s_1,s_2}{B(v)}
$
where $v \in  \set{[v_1;v_2],[v_2;v_1]}$.
Thus at the end of the instant, a dereferenced signal such as $!s_{1}$
becomes a {\em list} (possibly empty) 
of (distinct) values emitted on $s_1$ during the
instant and then all signals are reset. 

We continue our informal discussion with an example of 
a `server' handling a list of requests
emitted in the previous instant on the signal $s$. 
For each request of the shape $\s{req}(s',x)$, 
it provides an answer which is a function of $x$ along the signal $s'$
(the notation $x \unrhd p$  is used to match a value $x$ against a pattern $p$).
The `client' issues a request $x$ on signal $s$
and returns the reply on signal $t$.

{\footnotesize
\[
    \begin{array}{lcl}
      \w{Server}(s)&=&{\s{pause}}.\w{Handle}(s,!s)\\
      \w{Handle}(s,\ell)&=&
      \matchv{\ell}{\s{cons}(\s{req}(s',x),\ell')}
      {(\emit{s'}{f(x)} \mid \w{Handle}(s,\ell'))}
      {\w{Server}(s)} \\
      \w{Client}(x,s,t) &=&\nu s' \ (\emit{s}{\s{req}(s',x)} \mid \s{pause}.s'(x).\emit{t}{x},0)~.
    \end{array}
\]}
\noindent

Let us first notice that a request contains a `pointer', namely the
name of the signal on which to answer the request. Then the `folklore
solution' of transforming a list of values into one value via an
associative and commutative function does not work here. Indeed there
seems to be no reasonable way to define an associative and commutative
function on pointers. Instead, we look at \w{Handle} as
a function from (a signal and) a list of requests to behaviours
which is invariant under permutations of the list of requests.
Note that to express this invariance we need a notion of
behavioural equivalence and that this equivalence must 
satisfy the usual associativity and commutativity laws of
parallel composition and must be preserved by parallel composition.

These considerations are enough to argue that the \w{Server}
is a `deterministic' program. No matter how many clients
will issue requests at each instant, the \w{Server} will
provide an answer to each of them in the following instant
in a way which is independent of the order of the requests.
Let us now look at the \w{Client}. After issuing a request, the
\w{Client} waits for a reply in the following instant. Clearly, if
more than one reply comes, the outcome of the computation is not
deterministic. For instance, we could have several `Servers' running in
parallel or a server could somehow duplicate the request.  This means
that the usage of the signal $s$ must be such that many `clients' may
issue a request but at most one `server' may handle them at the end of
the instant in an `affine' way.  
Further, on the client side, the return signal $s'$ can
only be used to read while on the server side it can only be used to
emit.

This preliminary discussion suggests the need for a formal analysis of
the principles that allow to establish the determinacy of a
synchronous program. This analysis will be obviously inspired by previous
work on the foundations of linear logic \cite{Girard87}, on linear
typing of functional programs ({\em e.g.}, \cite{Wadler93}), and on
linear usages of channels ({\em e.g.}, \cite{KPT99}).  Following this
line of works, the analysis presented in section \ref{typing-sec} will
take the form of a {\em typing system}.  The previous section
\ref{section-spi}, will recall the formal definition of the
$\spi$-calculus.
In the final section \ref{results-sec},  
first we shall introduce the properties of the typing  system leading
to a {\em subject reduction} theorem, 
and second we shall describe a suitable notion of typed bisimulation and show
that with respect to this notion, typable programs can be regarded
as {\em deterministic}.

\section{Definition of the $\spi$-calculus}\label{section-spi}
We recall the formal definition of the $\spi$-calculus and its
bisimulation based semantics while referring the reader to 
\cite{Amadio06,AM07} for a deeper analysis.
This section is rather technical but to understand 
the type system described in the following section \ref{typing-sec}
there are really just two points that the reader should
keep in mind:

\begin{enumerate}
\item The semantics of the calculus is given by the labelled
  transition system presented in table \ref{Lts}.  A reader familiar
  with a $\pi$-calculus with asynchronous communication can understand
  these rules rather quickly.  The main differences are (a) the rule
  for emitting a signal formalises the fact that a signal, unlike a
  channel, persists within an instant and (b) the rules that describe
  the computation at the end of the instant.
\item The labelled transition system induces a rather standard notion
  of bisimulation equivalence (definition \ref{def-bis}) which is preserved by
  {\em static} contexts (fact \ref{bis-cong-fact}).\footnote{As a matter of
    fact the labelled transition system is built so that the definition of 
    bisimulation equivalence looks standard \cite{AM07}.} 
  In section \ref{results-sec}, we shall introduce a `typed' definition of the
  bisimulation and show that with respect to this definition, typable
  programs are deterministic.
\end{enumerate}

\subsection{Programs}\label{programs}
Programs $P,Q,\ldots$ in the $\spi$-calculus
are defined in table \ref{syntax}.
We use the notation $\vc{m}$ for a vector $m_1,\ldots,m_n$, $n\geq 0$.
The informal behaviour of programs follows.
$0$ is the terminated thread. $A(\vc{e})$ is a (tail) recursive call
of a thread identifier $A$ with a vector $\vc{e}$ of expressions as argument;
as usual the thread identifier $A$ is defined by a unique equation
$A(\vc{x})=P$ such that the free variables of $P$ occur in $\vc{x}$.
$\emit{s}{e}$ evaluates the expression $e$ and emits its value on the
signal $s$.  $\present{s}{x}{P}{K}$ is the {\em present} statement
which is the fundamental operator of the model \cite{Amadio05}. If the values
$v_1,\ldots,v_n$ have been emitted on the signal $s$ 
then $\present{s}{x}{P}{K}$ evolves
non-deterministically into $[v_i/x]P$ for some $v_i$ ($[\_/\_]$ is our
notation for substitution).  On the other
hand, if no value is emitted then the continuation $K$ is evaluated at
the end of the instant.  $\match{s_1}{s_2}{P_1}{P_2}$ is the usual
matching function of the $\pi$-calculus 
that runs $P_1$ if $s_1$ equals $s_2$ and $P_2$, otherwise.
Here both $s_1$ and $s_2$ are free.
$\matchv{u}{p}{P_1}{P_2}$, matches $u$ against the pattern $p$.
We assume $u$ is either a variable $x$ or a value $v$ and $p$ has
the shape $\s{c}(\vc{x})$, where $\s{c}$ is a constructor and 
$\vc{x}$ is a vector of distinct variables.  We also assume that if $u$ is a
variable $x$ then $x$ does not occur free in $P_{1}$.
At run time, $u$ is always a {\em value} and we 
run $\theta P_1$ if $\theta=\w{match}(u,p)$ is the 
substitution matching $u$ against $p$, and $P_2$ if the 
substitution does not exist (written $\w{match}(u,p)\ucl$).
Note that as usual the variables occurring in the pattern $p$
(including signal names) are bound in $P_1$.
$\new{s}{P}$ creates a new signal name $s$ and runs $P$.
$(P_1\mid P_2)$ runs in parallel $P_1$ and $P_2$.  A continuation $K$
is simply a recursive call whose arguments are either expressions
or values associated with signals at the end of the instant in
a sense that we explain below. We shall also write 
$\s{pause}.K$ for $\new{s}{\present{s}{x}{0}{K}}$ with $s$ not free in $K$. 
This is the program that waits till the end of the instant and then
evaluates $K$.

\begin{table}
{\footnotesize
\[
\begin{array}{lll}
P &::= 0 \Alt A(\vc{e}) \Alt \emit{s}{e} \Alt \present{s}{x}{P}{K}
\Alt &\mbox{(programs)} \\

 &\qquad\match{s_1}{s_2}{P_1}{P_2} \Alt \matchv{u}{p}{P_1}{P_2} \Alt \new{s}{P}
\Alt P_1\mid P_2 \\

K &::=A(\vc{r})     &\mbox{(continuation next instant)}\\

\w{Sig} &::= s \Alt t \Alt \cdots  &\mbox{(signal names)} \\
\w{Var} &::= \w{Sig} \Alt x \Alt y \Alt z \Alt \cdots   &\mbox{(variables)} \\
\w{Cnst} &::= \s{*} \Alt \s{nil} \Alt \s{cons} \Alt \s{c} \Alt \s{d} \Alt\cdots &\mbox{(constructors)} \\
\w{Val} &::= \w{Sig} \Alt \w{Cnst}(\w{Val},\ldots,\w{Val})
&\mbox{(values $v,v',\ldots$)}\\
\w{Pat} &::=  \w{Cnst}(\w{Var},\ldots,\w{Var})
&\mbox{(patterns $p,p',\ldots$)} \\
\w{Fun} &::=f \Alt g \Alt \cdots &\mbox{(first-order function
  symbols)} \\
\w{Exp} &::= \w{Var} \Alt \w{Cnst}(\w{Exp},\ldots,\w{Exp}) \Alt 
                          \w{Fun}(\w{Exp},\ldots,\w{Exp}) 
&\mbox{(expressions $e,e',\ldots$)} \\
\w{Rexp} &::= {!\w{Sig}} \Alt \w{Var} \Alt 
\w{Cnst}(\w{Rexp},\ldots,\w{Rexp}) \Alt \\ &\qquad \w{Fun}(\w{Rexp},\ldots,\w{Rexp})
&\mbox{(exp. with deref. $r,r',\ldots$)} 

\end{array}
\]}
\caption{Syntax of programs and expressions}\label{syntax}
\end{table}

\subsection{Expressions}\label{expressions}
Expressions are partitioned in several syntactic categories
as specified in table~\ref{syntax}.
As in the $\pi$-calculus, signal names stand both for
signal constants as generated by the $\nu$ operator and signal
variables as in the formal parameter of the present operator.
Variables $\w{Var}$ include signal names as well as variables of other
types.  Constructors $\w{Cnst}$ include $\s{*}$, $\s{nil}$, and $\s{cons}$.
Values $\w{Val}$ are terms built out of constructors and signal names.
Patterns $\w{Pat}$ are terms built out of constructors and variables
(including signal names).  
If $P, p$ are a program and a pattern then  we denote
with $\w{fn}(P), \w{fn}(p)$ the set of free signal names
occurring in them, respectively. We also use $\w{FV}(P), \w{FV}(p)$
to denote the set of free variables (including signal names).
We assume first-order function symbols $f,g,\ldots$  and
an evaluation relation $\eval$ such that for every
function symbol $f$ and values $v_1,\ldots,v_n$ of suitable type
there is a unique value $v$ such that $f(v_1,\ldots,v_n)\eval v$ and
$\w{fn}(v)\subseteq \Union_{i=1,\ldots,n}\w{fn}(v_{i})$.
Expressions $\w{Exp}$ are terms built out of variables, constructors,
and function symbols. The evaluation relation $\eval$ is extended
in a standard way to expressions whose only free variables are
signal names.
Finally, $\w{Rexp}$ are expressions that may include 
the value associated with a signal $s$ at the
end of the instant (which is written $!s$, following the ML notation
for dereferenciation). Intuitively, this value
is a {\em list of values} representing the set of values emitted on 
the signal during the instant.

The definition of a {\em simple} type system for the $\spi$-calculus
can be extracted from the more elaborate type system presented
in section \ref{typing-sec} by confusing `set-types' with `list-types'
and by neglecting all considerations on usages.


\subsection{Actions}\label{action-sec}
The syntactic category $\w{act}$ of {\em actions} described in table \ref{Lts}
comprises relevant, auxiliary, and nested actions.
The operations $\w{fn}$ (free names), $\w{bn}$ (bound names), 
and $\w{n}$ (both free  and bound names) are defined as in the
$\pi$-calculus \cite{MPW92}.

The {\em relevant actions} are those that are actually considered in the
bisimulation game. They consist of: (i) an internal action $\tau$, (ii)
an emission action $\outact{t}{s}{v}$ where it is assumed that the
signal names $\vc{t}$ are distinct, occur in $v$, and differ from $s$,
(iii) an input action $sv$, and (iv) an action $N$ (for {\em Next})
that marks the move from the current to the next instant.

The {\em auxiliary actions} consist of an input action $s?v$ which is
coupled with an emission action in order to compute a $\tau$ action and
an action $(E,V)$ which is just needed to compute an action $N$.
The latter is an action that can occur {\em exactly} when
the program cannot perform $\tau$ actions and it amounts to (i) collect
in lists the set of values emitted on every signal, (ii) to reset all
signals, and (iii) to initialise the continuation $K$
for each present statement of the shape $s(x).P,K$.

In order to formalise these three steps we need to introduce 
some notation.
Let $E$ vary over functions from signal names
to finite sets of values. Denote with $\emptyset$
the function that associates the empty set with every
signal name,  with $[M/s]$ the function that associates 
the set $M$ with the signal name $s$ and the empty set
with all the other signal names, and with $\union$ the 
union of functions defined point-wise.

We represent a set of values as a list of the
values belonging to the set. More precisely,
we write $v \real M$ and say that $v$ {\em represents} $M$ 
if $M=\set{v_1,\ldots,v_n}$ and
$v=[v_{\pi(1)};\ldots; v_{\pi(n)}]$ for
some permutation $\pi$ over $\set{1,\ldots,n}$.
Suppose $V$ is a function from signal names to lists of values.
We write $V\real E$ if $V(s)\real E(s)$ for every signal name $s$.
We also write $\w{dom}(V)$ for 
$\set{s \mid V(s)\neq []}$.
If $K$ is a continuation, {\em i.e.}, a recursive call $A(\vc{r})$,
then $V(K)$ is obtained from $K$ by replacing
each occurrence $!s$ of a dereferenced signal with the associated
value $V(s)$. We denote with $V[\ell/s]$ the function that behaves as
$V$ except on $s$ where $V[\ell/s](s)=\ell$.

With these conventions, a transition
$P\act{(E,V)} P'$ intuitively means that 
(1) $P$ is suspended,
(2) $P$ emits exactly the values specified by $E$, and
(3) the behaviour of $P$ in the following instant is $P'$ and depends
on  $V$. 
It is convenient to compute these transitions on programs where
all name generations are lifted at top level. We write $P \nor Q$
if we can obtain $Q$ from $P$ by repeatedly  transforming, for instance,
a subprogram $\nu s P' \mid P''$ into $\nu s (P'\mid P'')$ 
where $s\notin \w{fn}(P'')$.

Finally, the {\em nested actions} $\mu,\mu',\ldots$ are certain
actions (either relevant or auxiliary) that can be  produced by a sub-program 
and that we need to propagate to the top level. 
 
\subsection{Labelled transition system and bisimulation}
The labelled transition system 
is defined in table \ref{Lts} where rules apply to
programs whose only free variables are signal names and with
standard conventions on the renaming of bound names.
As usual, one can rename bound variables, and 
symmetric rules are omitted.
The first $12$ rules from $(\w{out})$ to $(\nu_{\w{ex}})$
are quite close to those of a polyadic $\pi$-calculus with
asynchronous communication (see \cite{HY95,ACS98})
with the following exception: rule $(\w{out})$ models the fact
that the emission of a value on a signal {\em persists} within
the instant. The last $5$ rules from $(0)$ to $(\w{next})$
are quite specific of the $\spi$-calculus
and determine how the computation is carried on at the end
of the instant (cf. discussion in \ref{action-sec}).

\begin{table}
{\footnotesize
\[
\begin{array}{|lll|}
\hline &&\\
\w{act} 
&::= \alpha \Alt \w{aux}              
&(\mbox{actions}) \\

\alpha  
&::= \tau \Alt \outact{t}{s}{v} \Alt sv \Alt N 
&(\mbox{relevant actions})\\

\w{aux} 
&::= s?v \Alt (E,V) 
&(\mbox{auxiliary actions})\\

\mu     
&::= \tau \Alt \outact{t}{s}{v} \Alt s?v 
&(\mbox{nested actions}) \\
&&\\ \hline  
\end{array}
\]
\[
\begin{array}{cc}
\\

(\w{out})\quad\infer{e\eval v}{\emit{s}{e} \act{\emit{s}{v}} \emit{s}{e}}

&(\w{in}_{\w{aux}})\quad
\infer{~}{\present{s}{x}{P}{K} \act{s?v} [v/x]P}
\\ \\

(\w{in})\quad \infer{~}{P\act{sv} (P \mid \emit{s}{v})} 

&(\w{rec})\quad \infer{A(\vc{x})=P,\quad \vc{e}\eval \vc{v}}{A(\vc{e})\act{\tau}[\vc{v}/\vc{x}]P}  \\ \\

(=_{1}^{\w{sig}})\quad \infer{~}{\match{s}{s}{P_1}{P_2}\act{\tau} P_1}

&(=_{2}^{\w{sig}})\quad 
\infer{s_1\neq s_2}{\match{s_1}{s_2}{P_1}{P_2}\act{\tau} P_2} \\ \\

(=_{1}^{\w{ind}})\quad
\infer{\w{match}(v,p)=\theta}{\matchv{v}{p}{P_1}{P_2}\act{\tau} \theta P_1}

&(=_{1}^{\w{ind}})\quad
\infer{\w{match}(v,p)=\ucl}{\matchv{v}{p}{P_1}{P_2} \act{\tau} P_2}  \\ \\

(\w{comp})\quad
\infer{P_1\act{\mu} P'_1 \quad \w{bn}(\mu)\inter\w{fn}(P_2)=\emptyset}
{P_1\mid P_2\act{\mu} P'_1\mid P_2}

&(\w{synch})\quad
\infer{\begin{array}{c}
P_1 \act{\outact{t}{s}{v}} P'_1\quad P_2 \act{s?v}P'_2\\
\set{\vc{t}}\inter \w{fn}(P_2)=\emptyset\end{array}}
{P_1\mid P_2 \act{\tau} \new{\vc{t}}{(P'_1\mid P'_2)}} \\ \\

(\nu)\quad \infer{P\act{\mu} P' \quad t\notin n(\mu)}
{\new{t}{P}\act{\mu} \new{t}{P'}}

&(\nu_{\w{ex}})\quad
\infer{P\act{\outact{t}{s}{v}} P'\quad t'\neq s\quad t'\in n(v)\minus \set{\vc{t}}}
{\new{t'}{P}\act{(\nu t',\vc{t})\emit{s}{v}} P'}  \\ \\

(0)\quad
\infer{~}
{0 \act{\emptyset,V} 0}

&(\w{reset})\quad
\infer{e\eval v \quad v \mbox{ occurs in }V(s)}
{\emit{s}{e} \act{[\set{v}/s],V} 0} \\ \\

(\w{cont})\quad
\infer{s\notin \w{dom}(V)}
{s(x).P,K \act{\emptyset,V} V(K)} 

&(\w{par})\quad
\infer{P_i \act{E_{i},V} P'_i \quad i=1,2 }
{(P_1\mid P_2) \act{E_1\union E_2,V} (P'_1\mid P'_2)} \\ \\

(\w{next})\quad
\infer{P\nor \nu \vc{s}\  P' \quad V\real E \quad P' \act{E,V} P''}
{P\act{N} \nu \vc{s} \ P''}

\end{array}
\]}
\caption{Labelled transition system}\label{Lts}
\end{table}

We derive from the labelled transition system a notion of (weak)
labelled bisimulation. First define $\wact{\alpha}$ as
$(\act{\tau})^*$ if $\alpha=\tau$,
$(\wact{\tau})\comp(\act{N})$ 
if $\alpha=N$, and 
$(\wact{\tau})\comp (\act{\alpha})\comp (\wact{\tau})$
otherwise.
This is the standard definition except that we insist on {\em not}
having internal reductions after an $N$ action.
Intuitively, we assume that an observer can control the execution
of programs so as to be able to test them
at the very beginning of each instant.
We write $P\act{\alpha} \cdot$ for $\xst{P'}{(P\act{\alpha}P')}$.

\begin{definition}[labelled bisimulation]\label{def-bis}
A symmetric relation $\rel{R}$ on programs is a labelled bisimulation
if $P\rl{R} Q$, $P\act{\alpha} P'$, $\w{bn}(\alpha)\inter
\w{fn}(Q)=\emptyset$ implies
$\xst{Q'}{( \ Q\wact{\alpha} Q',\quad P'\rl{R} Q' \ )}$.
We denote with $\wbis$ the largest labelled bisimulation.
\end{definition}


\begin{fact}[\cite{AM07}]\label{bis-cong-fact} 
Labelled bisimulation is preserved by parallel composition and name generation.
\end{fact}


\section{An affine type system}\label{typing-sec}
An analysis of the notion of determinacy carried on in \cite{AM07},
along the lines of \cite{Milner89}, suggests that
there are basically two situations that need to be analysed in order
to guarantee the determinacy of programs.  (1) At least two distinct
values compete to be received within an instant, for instance,
consider: $\emit{s}{v_{1}} \mid \emit{s}{v_{2}} \mid s(x).P,K$.  (2)
At the end of the instant, at least two distinct values are available
on a signal. For instance, consider: $\emit{s}{v_{1}} \mid
\emit{s}{v_{2}} \mid \s{pause}.A(!s)$.  A sensible approach is to
avoid completely the first situation and to allow the second provided
the behaviour of the continuation $A$ does not depend on the order in
which the values are collected. Technically, we consider a notion of
{\em affine signal usage} to guarantee the first condition and a
notion of {\em set type} for the second one.  While this is a good
starting point, it falls short of providing a completely satisfying
answer because the type constructions do {\em not} compose very well.
Then our goal is to discover a collection of {\em signal usages} with
better compositionality properties. The outcome of our analysis are
three new kinds of usages (kinds $3-5$ in table \ref{usage-table}).

\subsection{Usages}
In first approximation, we may regard a {\em usage} as an element of
the set $L=\set{0,1,\infty}$ with the intuition that $0$ corresponds
to no usage at all, $1$ to at most one usage, and $\infty$ to any
usage.  We {\em add} usages with a {\em partial} operation $\isum$
such that $0\isum a= a\isum 0 = a$ and $\infty\isum \infty=\infty$,
and which is undefined otherwise (note in particular that $1\isum 1$
is undefined).  The addition induces an {\em order} by $a\leq b$ if
$\xst{c}{a\isum c = b}$.  With respect to this order, $0$ is the least
element while $1$ and $\infty$ are {\em incomparable}.  If $a\geq b$
then we define a {\em subtraction} operation $a \ominus b$ as the {\em
largest} $c$ such that $a=b\oplus c$. Therefore: $a\ominus 0=a$,
$1\ominus 1 = 0$, and $\infty \ominus \infty = \infty$.

This classification of usages is adequate when handling purely
functional data where the intuition is that data with usage 1 have at
most one pointer to them \cite{Wadler93}.
However, when handling
more complex entities such as references, channels, or signals it is
convenient to take a more refined view.  Specifically, a usage
can be refined to include information about whether a signal is used:
(i) to emit, (ii) to receive during the instant, or (iii) to receive
at the end of the instant.  Then a usage becomes an element of $L^3$.
Among the 27 possible usages of the shape $(a,b,c)$ for $a,b,c\in L$,
we argue that there are $5$ {\em main} ones as described in table
\ref{usage-table} (left part). First of all, we must have $a\neq 0$
and $(b\neq 0 \Or c\neq 0)$ since a signal on which we cannot send or
receive has no interest.  Now if $a=\infty$ then we are forced to take
$b=0$ since we want to preserve the determinacy.  Then for $c=\infty$
we have the usage $e_1$ and for $c=1$ we have the usage $e_3$.
Suppose now $a=1$. One choice is to have $b=c=\infty$ and then we have
the usage $e_2$. On the other hand if we want to preserve affinity
then we should receive the emitted value at most once.  Hence we have
$b=0,c=1$ or $b=1,c=0$ which correspond to the usages $e_4$ and $e_5$,
respectively.  From these 5 {\em main} usages within an instant, we
obtain the {\em derived ones} (see again table \ref{usage-table}) by
simply turning one or more $1$'s to $0$'s. We only add, subtract,
compare usages in $L^3$ that are derived from the same main usage.

In a {\em synchronous} framework, it makes sense to consider how
usages vary over {\em time}.  The {\em simplest} solution would be to
look at signal usages of the shape $x^\omega$, $x\in L^3$, 
which are {\em invariant} under time.  However, to reason effectively on 
programs, we are led to consider signal usages of the shape $xy^\omega$ where
$x,y\in L^3$ are derived from the same main usage.

The reader may have noticed that in this discussion 
we have referred to increasingly complex
`usages' varying over $L$, $L^3$, and $(L^3)^\omega$. Henceforth
a signal usage belongs to $(L^3)^\omega$.
Usages are classified in 5 {\em kinds} as showed in table \ref{usage-table}.
\footnote{The fact that, {\em e.g.}, $(1,0,0)$ 
occurs both in the usages of kind 4 and 5 is a 
slight source of ambiguity which is resolved by
assuming that the kind of the usage is made explicit.}

We denote with $U$ the set of all these usages
and with $U(i)$ the set of usages of kind $i$, for $i=1,\ldots,5$.
We consider that the addition operation $\isum$ is defined only if 
$u,u'\in U(i)$ and $u\isum u'\in U(i)$ for some $i\in
\set{1,\ldots,5}$.
Similar conventions apply when comparing and subtracting usages.
If $u\in U$ then $\uparrow u$, the {\em shift} of $u$,
is the infinite word in $U$ obtained from $u$ by removing the first
character. This operation is always defined. If $u$ is a signal usage, 
then $u(i)$ for $i\geq 0$ 
denotes its $i^{\w{th}}$ character and
$u(i)_{j}$ for $j\in\set{1,2,3}$ the $j^{\w{th}}$ component of $u(i)$.

\begin{table}
{\footnotesize
\[
\begin{array}{l||r}

\begin{array}{l|l}

\mbox{main usages}           &\mbox{derived usages}  \\ \hline
e_1=(\infty,0,\infty) &\_ \\

e_2=(1,\infty,\infty)      &(0,\infty,\infty) \\

e_3=(\infty,0,1)      &(\infty,0,0) \\

e_4=(1,0,1)           &(1,0,0),(0,0,1),(0,0,0) \\

e_5=(1,1,0)           &(1,0,0),(0,1,0),(0,0,0) 

\end{array}

&
\begin{array}{c|c|c|c}

xy^\omega\in U(i) \mbox{ is}&\mbox{affine} &\mbox{uniform} &\mbox{aff. preserving} \\ \hline
i=1    &no               &yes              &no \\
i=2    &yes/no           &yes/no            &no \\
i=3    &yes/no           &yes/no            &yes \\
i=4    &yes/no           &yes/no            &yes \\
i=5    &yes/no           &yes/no            &yes 
\end{array}
\end{array}
\]
}

\caption{Usages and their classification}\label{usage-table}
\end{table}

We classify the usages according to 3 properties: 
affinity, uniformity, and preservation
of affinity. We say that a usage is {\em affine} if it contains a
$`1'$  and {\em
non-affine} otherwise. We also say that it is {\em uniform} if it is
of the shape $x^\omega$ and that it is {\em neutral} if it is the
neutral element with respect to the addition $\isum$ on the set of
usages $U(i)$ to which it belongs.  It turns out that the non-affine
signal usages are always uniform and moreover they coincide with the neutral
ones.  Finally, by definition, the usages in 
the sets $U(i)$ for $i=3,4,5$ are {\em affine
preserving} 
The classification is summarised in the table \ref{usage-table} 
(right part).

\subsection{Types}
In first approximation, types are either {\em inductive types} or {\em
signal} types. As usual, an inductive type such as the type
$\w{List}(\sigma)$ of lists of elements of type $\sigma$ is defined by
an equation $\w{List}(\sigma) = \s{nil} \Alt \s{cons} \ \w{of} \
\sigma,\w{List}(\sigma)$ specifying the ways in which an element of
this type can be built.

In our context, inductive types come with a 
{\em usage} $x$ which belongs to the set  $\set{1,\infty}$ and 
which intuitively specifies whether the values of this type
can be used at most once or arbitrarily many times (once more we
recall that $1$ and $\infty$ are incomparable).
To summarise, if $\sigma_1,\ldots,\sigma_k$ are types already defined
then an inductive type $C_{x}(\sigma_1,\ldots,\sigma_k)$ is defined 
by case on constructors of the shape 
{\small
$\s{c} \ \w{of} \ \sigma'_1,\ldots,\sigma'_m$}
where the types $\sigma'_j$, $j=1,\ldots,m$ 
are either one of the types $\sigma_i$, $i=1,\ldots,n$ or the
inductive type $C_{x}(\ldots)$ being defined.  
There is a further constraint that has to be respected, 
namely that if one of the types $\sigma_i$ is `affine' then
the usage $x$ must be affine preserving, {\em i.e.}, $x=1$.
An affine type is simply a type which contains an affine usage.
The grammar in table \ref{aff-type-sys} will provide a precise definition
of the affine types.

When collecting the values at the end of the instant
we shall also need to consider  {\em set types}.
They are described by an equation
$\w{Set}_{x}(\sigma) = \s{nil} \Alt \s{cons} \ \w{of} \ \sigma,\w{Set}_{x}(\sigma)$
which is quite similar to the one for lists. Note that set types too
come with a usage $x\in\set{1,\infty}$ and that if $\sigma$ is an
affine type then the usage $x$ must be affine preserving.
The reader might have noticed that we take the freedom of using 
the constructor $\s{nil}$ both with the types
$\w{List}_{u}(\sigma)$ and $\w{Set}_{u}(\sigma)$, $u\in \set{1,\infty}$,
and the constructor $\s{cons}$ both with the types 
$(\sigma,\w{List}_{u}(\sigma))\arrow \w{List}_{u}(\sigma)$
and $(\sigma,\w{Set}_{u}(\sigma))\arrow \w{Set}_{u}(\sigma)$.
However, one should assume that a suitable label 
on the constructors will allow to disambiguate the situation.

Finally, we denote with $\w{Sig}_{u}(\sigma)$ the type of signals
carrying values of type $\sigma$ according to the signal usage $u$.
As for inductive and set types, if $\sigma$ is an affine
type then the signal usage $u$ must be affine preserving.
To formalise these distinctions, we are lead to use several names
for types as specified in table \ref{aff-type-sys}.  
We denote with $\kappa$ non-affine (or classical) types, 
{\em i.e.}, types  that carry {\em no} affine information. 
These types have a uniform usage.
We denote with $\lambda$ affine and uniform types.
The types $\sigma,\sigma',\ldots$ stand for types with uniform usage
(either non-affine or affine).  Finally, the types $\rho,\rho',\ldots$
include all the previous ones plus types that have a non-uniform usage.
We notice that classical uniform types can be nested in an arbitrary way,
while affine uniform types can only be nested under type constructors
that preserve affinity. Moreover, types with non-uniform usages (either
classical or affine) cannot be nested at all.\footnote{What's the meaning of 
sending a data structure containing informations whose usage is 
time-dependent? Is the time information relative to the instant 
where the data structure is sent or used? 
We leave open the problem of developing
a type theory with usages more complex than the ones of the shape
$xy^\omega$ considered here.}

The partial operation of addition $\isum$ is extended to types
so that:
$\w{Op}_{u_{1}}(\sigma) \oplus \w{Op}_{u_{2}}(\sigma) = 
\w{Op}_{u_{1}\oplus u_{2}}(\sigma)$,
where $Op$ can be $C$, $\w{Set}$, or $\w{Sig}$, and
provided that $u_{1}\oplus u_{2}$ is defined.
For instance, $\w{List}_{1}(\lambda)\oplus \w{List}_{1}(\lambda)$
is undefined because $1\oplus 1$ is not defined.

A type context (or simply a context) $\Gamma$ is a 
partial function with finite domain $\w{dom}(\Gamma)$
from variables to types.
An addition operation $\Gamma_1\isum \Gamma_2$ on contexts is 
defined, written $(\Gamma_1\isum \Gamma_2)\dcl$, if and only if for all $x$ such that $\Gamma_1(x)=\rho_1$ and
$\Gamma_2(x)=\rho_2$, the type $\rho_1\isum\rho_2$ is defined.
The shift operation is extended to contexts so that $(\uparrow
\Gamma)(x) = \w{Sig}_{(\uparrow u)}(\sigma)$ if
$\Gamma(x)=\w{Sig}_{u}(\sigma)$ and $(\uparrow \Gamma)(x) = \Gamma(x)$
otherwise.  We also denote with $\Gamma,x:\sigma$ the context $\Gamma$
{\em extended} with the pair $x:\sigma$ (so $x\notin
\w{dom}(\Gamma)$).  We say that a context is {\em neutral} ({\em
uniform}) if it assigns to variables neutral (uniform) types.

\subsection{Semantic instrumentation}
As we have seen, each signal belongs to exactly one of 5 kinds of
usages. Let us consider in particular the kind 5 whose main usage is
$e_{5}$.  The forthcoming type system  is
supposed to guarantee that a value emitted on a signal of kind 5 is received at
most once during an instant.  Now, consider the program $\ol{s}t
\mid s(x).\ol{x},0$ and attribute a usage $e_5^{\omega}$ to the signals $s$
and $t$. According to this usage this program should be well
typed. However, if we apply the labelled transition system in table
\ref{Lts}, this program reduces to $(\ol{s}t \mid \ol{t})$ which fails
to be well-typed because the double occurrence of $t$ is not
compatible with an affine usage of $t$.  Intuitively, after the signal
$s$ has been read once no other synchronisation should arise during
the instant either within the program or with the environment.
To express this fact we proceed as follows.
First, we instrument the semantics so that it marks (underlines) 
the emissions on signals of kind 5 that have been used at least once during the
instant. The emission has no effect on the labelled transition
system in the sense that $\ul{\emit{s}{e}}$ behaves exactly as $\emit{s}{e}$.

{\footnotesize \[
\begin{array}{ccc}

(\w{out})\quad\infer{e\eval v}{\emit{s}{e} \act{\emit{s}{v}} \ul{\emit{s}{e}}}\qquad

&(\w{\ul{out}})\quad\infer{e\eval v}{\ul{\emit{s}{e}} \act{\emit{s}{v}} \ul{\emit{s}{e}}} \qquad

&(\w{\ul{reset}})\quad
\infer{e\eval v \quad v \mbox{ occurs in }V(s)}
{\ul{\emit{s}{e}} \act{[\set{v}/s],V} 0} 

\end{array}
\]}

On the other hand, we introduce a special rule $(\ul{\w{out}})$ to type
$\ul{\emit{s}{e}}$ which requires at least a  usage 
$(1,1,0)\cdot (0,0,0)^\omega$ for the signal $s$ while neglecting the
expression $e$. By doing this, we make sure that a second attempt
to receive on $s$ will produce a type error. In other terms, if 
typing is preserved by `compatible' transitions, then we can be sure
that a value emitted on a signal of kind 5 is received at most once
within an instant.

\subsection{Type system}
The type system is built around few basic ideas.
(1) Usages including both input and output capabilities can be decomposed in
simpler ones. For instance,
$(1,1,0)^\omega = (1,0,0)(0,1,0)^\omega \isum (0,1,0)(1,0,0)^\omega$.
(2) A rely-guarantee kind of reasoning: when we emit a value 
we {\em guarantee} certain resources while when we receive a value
we {\em rely} on certain resources.
(3) Every affine usage can be consumed at most once in the typing judgement
(and in the computation).

When formalising the typing judgements we need to distinguish
the typing of an expression $e$ from the typing of an expression
with dereferenciation $r$ and the typing of a recursive call 
$A(e_1,\ldots,e_n)$ from the typing of a recursive call at the end of
the instant $A(r_1,\ldots,r_n)$. 
To do this we shall write $[r]$ rather than $r$ and 
$[A(r_1,\ldots,r_n)]$ rather than $A(r_1,\ldots,r_n)$.

We shall consider {\em four typing judgements}:
$\Gamma\Gives e:\rho$,
$\Gamma \Gives [r]:\rho$,
$\Gamma \Gives P$,
and
$\Gamma \Gives [A(r_1,\ldots,r_n)]$,
and we wish to refer to them
with a {\em uniform} notation $\Gamma \Gives U:T$.
To this end, we introduce a  fictious type $\w{Pr}$ of programs and regard the judgements
$\Gamma \Gives P:\w{Pr}$ and $\Gamma \Gives  [A(r_1,\ldots,r_n)]:\w{Pr}$
as an expansion of $\Gamma \Gives P$ and $\Gamma \Gives [A(r_1,\ldots,r_n)]$, 
respectively. 
Then we let $U$ stand for one of $e$, $[r]$, $P$, $[A(r_1,\ldots,r_n)]$,
and $T$ for one of $\rho,\w{Pr}$.

We assume that function symbols are given 
non-affine types of the shape $(\kappa_1,\ldots,\kappa_n)\arrow \kappa$.
We denote with $k$ either a constructor or a function symbol and
we assume that its type is explicitly given. 

\begin{table}
{\footnotesize
\[
\begin{array}{|@{\quad}lll@{\quad}|}
\hline &&\\
\kappa&::= C_\infty(\vc{\bf \kappa}) \Alt \w{Set}_{\infty}(\kappa) 
\Alt \w{Sig}_{u}(\kappa) 
&(u \mbox{ neutral}) \\ 

\lambda&::= C_1(\vc{\bf \sigma}) \Alt \w{Set}_{1}(\sigma) \Alt 
\w{Sig}_{u}(\kappa) \Alt \w{Sig}_{v}(\lambda) 
&(u \mbox{ affine and uniform}, v \mbox{ aff.-pres.} \\ 
&&\mbox{ and uniform)}\\

\sigma&::=\kappa \Alt \lambda 
&(\mbox{uniform types})\\

\rho&::=\sigma \Alt \w{Sig}_{u}(\kappa) \Alt \w{Sig}_{v}(\lambda)
&(v \mbox{ affine-preserving})  \\
&&\\\hline

\end{array}
\]
\[
\begin{array}{cc}

(\w{var})\quad
\infer{u\geq u'\quad \w{Op}\in\set{\w{Sig},\w{Set},C}}
{\Gamma,x:\w{Op}_{u}(\vc{\sigma}) \Gives x: \w{Op}_{u'}(\vc{\sigma})}  

&(k)\quad
  \infer{\begin{array}{c}
      \Gamma_i \Gives e_i:\sigma_i\quad i=1,\ldots,n
      \\ 
      k:(\sigma_1,\ldots,\sigma_n)\arrow \sigma \quad k=f \mbox{ or }k=\s{c}
    \end{array}}
  {\Gamma_0\isum \Gamma_1\isum \cdots \isum \Gamma_n \Gives
    k(e_1,\ldots,e_n): \sigma} \\ \\

[\w{var}_{C}]\quad
\infer{\w{Op}=\w{C}\quad \w{Op}=\w{Set}}
{\Gamma,x:\w{Op}_{u}(\sigma) \Gives [x]: \w{Op}_{u}(\sigma)}  

&[\w{var}_{\w{sig}}] \quad
\infer{y^\omega \geq u}
{\Gamma,s:\w{Sig}_{xy^\omega}(\sigma) \Gives [s]: \w{Sig}_{u}(\sigma)}  \\ \\ 

\multicolumn{2}{c}{[k]\quad
\infer{\begin{array}{c}
        \Gamma_i \Gives [ r_i ] :\sigma_i\quad i=1,\ldots,n
	\\ 
        k:(\sigma_1,\ldots,\sigma_n)\arrow \sigma \quad k=f \mbox{ or }k=\s{c}
        \end{array}}
{\Gamma_0\isum \Gamma_1\isum \cdots \isum \Gamma_n \Gives [k(r_1,\ldots,r_n)]: \sigma}}  \\ \\

[!_{\w{Set}}]\quad
\infer{\begin{array}{c}
(u(0) \geq (\infty,0,\infty)\  \AND  \ x= \infty) \ \Or \\
(u(0) \geq (\infty,0,1) \ \AND \  x= 1) \\
\end{array}}
{\Gamma,s:\w{Sig}_{u}(\sigma)  \Gives [!s]:\w{Set}_{x}(\sigma)} 

&[!_{\w{List}}]\quad
\infer{\begin{array}{c}
(u(0)\geq (0,\infty,\infty) \ \AND \ x= \infty) \ \Or \\
(u(0)\geq (0,0,1)  \ \AND \  x= 1)   
\end{array}}
{\Gamma,s:\w{Sig}_{u}(\sigma)  \Gives [!s]:\w{List}_{x}(\sigma)}   \\
\\ 

(0)\quad
\infer{~}{\Gamma \Gives 0}

&(\w{out})\quad
\infer{
\begin{array}{c}
\Gamma_1 \Gives s:\w{Sig}_{u}(\sigma) \quad u(0)_{1}\neq 0 \\
\Gamma_2 \Gives e:\sigma
\end{array}
}
{\Gamma_1 \isum \Gamma_2  \Gives \emit{s}{e}} \\ \\

(\nu)\quad
\infer{ 
\begin{array}{c}
\Gamma,s:\w{Sig}_{u}(\sigma)\Gives P
\end{array}}
{\Gamma \Gives \nu s:\w{Sig}_{u}(\sigma) \ P}

&
(\w{in})
\quad
\infer{
\begin{array}{c}
\Gamma_1 \Gives s: \w{Sig}_{u}(\sigma) \quad u(0)_{2}\neq 0 \\
\Gamma_2,x:\sigma \Gives P\quad (\Gamma_1\isum \Gamma_2) \Gives [A(\vc{r})]
\end{array}}
{(\Gamma_1\isum \Gamma_2) \Gives s(x).P,A(\vc{r}) } \\ \\

(m_s)\quad
\infer{\begin{array}{c}
s_1,s_2\in \w{dom}(\Gamma) \\
\Gamma \Gives P_i\quad i=1,2
\end{array}}
{\Gamma \Gives [s_1=s_2]P_1,P_2} 

&(m_{\s{c}})\quad
\infer{
\begin{array}{c}
\s{c}:(\sigma_1,\ldots,\sigma_n)\arrow \sigma 
\quad \Gamma_1 \Gives u:\sigma \\
\Gamma_2,x_1:\sigma_1,\ldots,x_n:\sigma_n \Gives P_1
\\ (\Gamma_1\isum \Gamma_2) \Gives P_2
\end{array}}
{\Gamma_1 \isum \Gamma_2 \Gives \matchv{u}{\s{c}(x_1,\ldots,x_n)}{P_1}{P_2}}   \\ \\

(\w{par})
\quad
\infer{\Gamma_i\Gives P_i\quad i=1,2}
{\Gamma_1\isum \Gamma_2 \Gives P_1\mid P_2}

&(\w{rec})
\quad
\infer{\begin{array}{c}
        A:(\sigma_1,\ldots,\sigma_n),\\
       \Gamma_i \Gives e_i:\sigma_i\quad i=1,\ldots,n
       \end{array}}
{\Gamma_1\isum \cdots \isum \Gamma_n \Gives A(e_1,\ldots,e_n)}  \\ \\

(\w{\ul{out}})
\quad
\infer{
\Gamma \Gives s: \w{Sig}_{u}(\sigma) \quad 
u(0)=(1,1,0)
}
{\Gamma \Gives \ul{\emit{s}{e}} }

&[\w{rec}]
\quad
\infer{\begin{array}{c}
        A:(\sigma_1,\ldots,\sigma_n),\\
       \Gamma_i \Gives [r_i]:\sigma_i\quad i=1,\ldots,n
       \end{array}}
{\Gamma_1\isum \cdots \isum \Gamma_n \Gives [A(r_1,\ldots,r_n)]}  

\end{array}
\]}
\caption{Affine type system}\label{aff-type-sys}
\end{table}

The typing rules are given in table \ref{aff-type-sys}.
We comment first on the typing rules for the expressions.
We notice that the arguments and the result of a constructor or a function
symbol have always a uniform type.  The rules $(!_{\w{Set}})$ and 
$(!_{\w{List}})$ describe the
type of a dereferenced signal following its usage. If the usage
is of kind 1 then the list of values associated with the signal at
the end of the instant must be treated as a set, if the usage
is of kind 2 then we know that the list of values contains
at most one element and therefore its processing will
certainly be `order-independent', if the usage is  of kind 3
then the list may contain several values and it must be processed as
an {\em affine} set, finally if the usage is of kind 4 (the usage of
kind 5 forbids reception at the end of the instant)
then  again the list of values will contain at most one element so we
can rely on an {\em affine} list type.

Notice the special form of the rule $[\w{var}_{\w{sig}}]$.
The point here is that in a recursive call $K=A(!s,s)$ at the end of
instant, we need to distinguish the resources needed to type 
$!s$ which should relate to the {\em current} instant
from the resources needed to type $s$ which should relate to the {\em following
instants}. For instance, we want to type $K$ in a context
$s:\w{Sig}_{u}(\sigma)$ where $u=(0,0,1)^\omega$. This is possible
because we can decompose $u$ in $u_1\oplus u_2$, where 
$u_1=(0,0,1)(0,0,0)^\omega$ and $u_2=(0,0,0)(0,0,1)^\omega$, and we can 
rely on $u_1$ to type $[!s]$ and  on $u_2$ to type $[s]$ 
(by  $[\w{var}_{\w{sig}}]$).

A set-type is a particular case of quotient type and therefore
its definition goes through the definition of an 
equivalence relation $\sim_\rho$ on values. This is defined as the least 
equivalence relation such that $s \sim_{\w{Sig}_{u}(\sigma)} s$, 
$\s{c} \sim_{C(\vc{\sigma})} \s{c}$, if $\s{c}$ is a constant of type
$C(\vc{\sigma})$, and 

{\footnotesize
\[
\begin{array}{ll}
\s{c}(v_1,\ldots,v_n) \sim_{C_{u}(\sigma_1,\ldots,\sigma_n)} 
\s{c}(u_1,\ldots,u_n)
&\mbox{if }v_i \sim_{\sigma_{i}} u_i \mbox{ for }  
i=1,\ldots,n
\\

[v_1;\ldots;v_n] \sim_{\w{Set}_{u}(\sigma)} [u_1;\ldots;u_m] 
&\mbox{if }
\set{v_1,\ldots,v_n} \sim_{\w{Set}_{u}(\sigma)} \set{u_1,\ldots,u_m}, \\

\mbox{where: }
\set{v_1,\ldots,v_n} \sim_{\w{Set}_{u}(\sigma)} \set{u_1,\ldots,u_m}
&\mbox{if for a permutation }\pi, v_i \sim_{\sigma} u_{\pi(i)}~.

\end{array}
\]}

Furthermore, 
we assume that each function symbol $f$, coming with a (classical) type
$(\kappa_1,\ldots,\kappa_n)\arrow \kappa$,
{\em respects} the typing in the  following sense:
(1)  if $v_i \sim_{\kappa_{i}} u_i$, $i=1,\ldots,n$,
$f(v_1,\ldots,v_n)\eval v$ and $f(u_1,\ldots,u_n)\eval u$ then
$v\sim_{\kappa} u$.
(2) If $\Gamma \Gives f(v_1,\ldots,v_n):\kappa$
and $f(v_1,\ldots,v_n) \eval v$ then $\Gamma \Gives v:\kappa$.

Finally, we turn to the typing of programs.
We assume that each thread identifier $A$, 
defined by an equation  $A(x_1,\ldots,x_n)=P$,
comes with a type  $(\sigma_1,\ldots,\sigma_n)$.
Hence we require these types to be uniform.
We also require that $A$ has the property that: (i) 
if $v_i \sim_{\sigma_{i}} u_i$ for $i=1,\ldots,n$ then
$A(v_1,\ldots,v_n) \wbis A(u_1,\ldots,u_n)$
and   (ii) $x_1:\sigma_1,\ldots,x_n:\sigma_n \Gives P$ is derivable.

We also suppose that generated signals names are {\em explicitly} labelled with their
types as in $\nu s:\rho \ P$. The labelled transition system in table \ref{Lts}
is adapted so that the output action carries the information on the types of the
extruded names. This type is lifted by the rule $(\w{next})$ so that,
{\em e.g.}, $\nu s:\rho \ s.0,A(s) \act{N} \nu s: \uparrow \rho \ A(s)$.

\begin{example}
With reference to the example of client-server in section
\ref{intro-sec}, 
assume an inductive (non-affine) type $D$ of data.
Let $\sigma_{1} = \w{Sig}_{u_{1}}(D)$ where $u_{1}=(1,0,0)^\omega$  be the type of 
the signals on which the server will eventually provide an answer.
Let $\w{Req}_1(\sigma_1,D) = \s{req} \ \w{of} \ \sigma_r,D$ be
the type of requests which are pairs composed of a signal and a datum.
Let $\sigma_{\w{set}} = \w{Set}_{1}(\w{Req}_{1}(\sigma_1,D))$ be the type of the set
of requests issued by the clients.
Let $\sigma=\w{Sig}_{u}(\w{Req}_{1}(\sigma_1,D))$ 
with $u=(\infty,0,1)^\omega$ be the type
of the signal on which the server gets the requests
and $\sigma'= \w{Sig}_{u'}(\w{Req}_{1}(\sigma_1,D))$, 
with $u'=(\infty,0,0)^\omega$, the related type of the signal on 
which the clients send the requests.
Finally, let $\sigma_t = \w{Sig}_{u}(D)$ be the type of the signal
on which the client sends the received answer (with a suitable usage $u$).
Then we can type \w{Server} and \w{Client} as follows:
$\w{Server}:(\sigma)$, $\w{Handle}:(\sigma, \sigma_\w{set})$, and
$\w{Client}:(D,\sigma', \sigma_t)$.
\end{example}

\begin{remark}
In a practical implementation of the type system, one can expect
the programmer to assign a kind $(1-5)$ to each signal
and let the system infer a {\em minimum} usage which is compatible
with the operations performed by the program.
\end{remark}


\section{Results}\label{results-sec}
We start by stating the expected {\em weakening} and 
{\em substitution} properties of the type system. 

\begin{lemma}[weakening]\label{weak-lemma}
If $\Gamma\Gives U:T$ and $(\Gamma\oplus \Gamma')\dcl$ then
$(\Gamma\oplus \Gamma')\Gives U:T$.
\end{lemma}

\begin{lemma}[substitution]\label{sub-lemma}
If $\Gamma,x:\rho \Gives U:T$, $\Gamma'\Gives v:\rho$, and 
$(\Gamma\oplus \Gamma')\dcl$ then $(\Gamma\oplus \Gamma')\Gives [v/x]U:T$.
\end{lemma}

Next we specify when a context $\Gamma$ is {\em compatible} with
an action $\w{act}$, written $(\Gamma,\w{act})\dcl$.
Recall that $V$ and $E$ denote a function from signals to finite lists
of distinct values and finite sets of values, respectively.  If
$V(s)=[v_1;\ldots;v_n]$ then let $(V\minus E)(s)= \set{v_1,\ldots,v_n}\minus
E(s)$.  Then define a program $P_{(V\minus E)}$ as the parallel
composition of emissions $\emit{s}{v}$ such that $v\in (V\minus
E)(s)$. Intuitively, this is the emission on an appropriate signal of
all the values which are in $V$ but not in $E$. We also let $P_V$
stand for $P_{(V\minus \emptyset)}$ where $\emptyset(s)=\emptyset$ for
every signal $s$.

\begin{definition}
With each action $\w{act}$, we associate a  {\em minimal}  program 
$P_{\w{act}}$ that allows the action to take place:

{\footnotesize
\[
P_{\w{act}} = 
    \left\{ \begin{array}{ll}
    0 & \mbox{if } \w{act}=\tau \mbox{ or } \w{act}=N \\
    \emit{s}{v} & \mbox{if } \w{act}=sv \mbox{ or } \w{act}=s?v \\
    s(x).0,0 & \mbox{if } \w{act}=\emit{s}{v} \\
    P_{V\minus E} & \mbox{if } \w{act}=(E,V)
\end{array} \right.
\]}

\end{definition}

\begin{definition}[compatibility context and action]
A context $\Gamma$ is compatible with an action 
$\w{act}$, written $(\Gamma,\w{act})\dcl$, if 
$\xst{\Gamma'}
{(\Gamma \oplus \Gamma')\dcl \mbox{ and }
\Gamma' \Gives P_{act}}$.
\end{definition}

We can now introduce the concept of  {\em typed} transition 
which is a transition labelled with an action $\w{act}$ 
of a program  typable in a context $\Gamma$ 
such that $\Gamma$ and $\w{act}$ are compatible.

\begin{definition}[typed transition]
We write $P\tact{\Gamma}{\w{act}} Q$ ($P\twact{\Gamma}{\w{act}} Q$)
if:
(1) $\Gamma \Gives P$, 
(2) $(\Gamma,\w{act})\dcl$, and (3)
$P\act{\w{act}} Q$ ($P\wact{\w{act}} Q$, respectively). 

\end{definition}

Next, we introduce the notion of {\em residual context} which
is intuitively the context left after a typed transition. 
(the definition for the auxiliary actions is available in 
 appendix \ref{residual-aux-sec}).
First, we notice that given a (uniform) type $\sigma$ and a value
$v$ we can define the minimum context $\Delta(v,\sigma)$ such that
$\Delta(v,\sigma)\Gives v:\sigma$. Namely, we set
$\Delta(s,\sigma)  = s:\sigma$ and
$\Delta(\s{c}(v_1,\ldots,v_n)) = 
\Delta(v_1,\sigma_1)\oplus \cdots \oplus \Delta(v_n,\sigma_n)$
 if $c:(\sigma_1,\ldots,\sigma_n)\arrow \sigma$.
Notice that $\Delta(v,\sigma)$ is the empty context if
$\w{fn}(v)=\emptyset$ and it is a neutral context if $\sigma$ is
non-affine.

\begin{definition}[residual context]\label{res-cxt-def}
Given a context $\Gamma$ and a compatible and relevant action $\w{\alpha}$, the
residual context $\Gamma(\w{\alpha})$ is defined as follows:

{\footnotesize
  \[
  \Gamma(\alpha) = \left\{
    \begin{array}{ll}
      \Gamma 
      & \mbox{if } \alpha=\tau \\

      \ucl\Gamma 
      & \mbox{if }\alpha=N \\

      (\Gamma,\vc{t:\sigma'})\ominus\Delta(v:\sigma')\oplus\set{s:Sig_{u_5}(\sigma')}
      & \mbox{if }\Gamma(s) = Sig_u(\sigma'), \alpha=\nu
      \vc{t:\sigma'}\emit{s}{v}, (1)\\
     
      \Gamma\oplus\Delta(v,\sigma')\oplus\set{s:Sig_{u_{out}}(\sigma')}
      & \mbox{if }\Gamma(s) = Sig_u(\sigma'), \alpha=sv, (2)

    \end{array}
  \right.
  \]}

\noindent (1)  $u_5=\usage{0}{1}{0}\cdot\om{\usage{0}{0}{0}}$ if $u\in
  U(5)$ and it is neutral otherwise ({\em i.e.}, $u\in U(2)$).
(2) $u_{out}$ is the least usage of the same
kind  as $u$ which allows to perform an output within the instant 
(always defined).
\end{definition}

The notion of residual context is instrumental to a precise statement
of the way transitions affect the typing. First we notice that the
type of expressions is preserved by the evaluation relation.

\begin{lemma}[expression evaluation]\label{exp-eval-lemma}
If $\Gamma \Gives e:\rho$ and $e\eval v$ then $\Gamma \Gives v:\rho$.
\end{lemma}

The following lemma records the effect of the substitution
at the end of the instant.

\begin{lemma}[substitution, end of instant]\label{sub-eoi}
\Defitemf{(1)}
If $\Gamma \Gives [A(\vc{r})]$, $\Gamma'\Gives P_V$, and 
$(\Gamma \oplus \Gamma')\dcl$ then $\uparrow (\Gamma \oplus \Gamma')\Gives
V(A(\vc{r}))$. 

\Defitem{(2)} If moreover there are $V',E$ such that $V,V'\real E$
then $V(A(\vc{r}))\wbis V'(A(\vc{r}))$.
\end{lemma}

Finally, the subject reduction theorem states that the residual 
of a typed transition is typable in the residual context 
(again, the residual context on auxiliary actions is defined in
appendix \ref{residual-aux-sec}).

\begin{theorem}[subject reduction]\label{subj-red-thm}
If $P \tact{\Gamma}{\w{act}} Q$ then
$\Gamma(\w{act}) \Gives Q$.
\end{theorem}

Next we introduce a notion of {\em typed bisimulation} which refines the one
given in definition \ref{def-bis} by focusing on typed processes and 
typed transitions. Let $\w{Cxt}$ be the set of contexts and if $\Gamma\in \w{Cxt}$
let $\w{Pr}(\Gamma)$ be the set of programs typable in the context
$\Gamma$.

\begin{definition}[typed bisimulation]
A typed bisimulation is a function $\rl{R}$ indexed on
$\w{Cxt}$ such that for every context $\Gamma$, 
$\trel{R}{\Gamma}$ is a symmetric relation on $\w{Pr}(\Gamma)$  such that:
$P \trl{R}{\Gamma} Q$, $P\tact{\Gamma}{\alpha} P'$, 
$\w{bn}(\alpha)\inter \w{fn}(Q)=\emptyset$ implies
$\xst{Q'}{( \ Q\twact{\Gamma}{\alpha} Q',\quad P'\trl{R}{\Gamma(\alpha)} Q' \ )}$.
We denote with $\twbis$ the largest typed labelled bisimulation.
\end{definition}

An expected property of typed bisimulation is that it is a weaker
property than untyped bisimulation: if we cannot distinguish two processes
by doing arbitrary actions we cannot distinguish them when 
doing actions which are compatible with the typing.

\begin{proposition}\label{free-typed-bis}
If $P,Q\in \w{Pr}(\Gamma)$ and  $P\wbis Q$ then $P\twbis_{\Gamma} Q$.
\end{proposition}

We write $P \tacteq{\Gamma}{\tau} Q$ if $P\tact{\Gamma}{\tau} Q$ or
$P=Q$.  The following lemma states a strong commutation property of
typed $\tau$ actions and it entails that typed bisimulation is
invariant under $\tau$-actions. 

\begin{lemma}\label{key-lemma1}
\Defitemf{(1)} 
If $P\tact{\Gamma}{\tau} P_i$ for $i=1,2$ then there is a $Q$
such $P_i \tacteq{\Gamma}{\tau} Q$ for $i=1,2$.

\Defitem{(2)} If $P\twact{\Gamma}{\tau} Q$ then $P\twbis_{\Gamma} Q$.
\end{lemma}

The second key property is that the computation at the end of the
instant is deterministic and combining the two lemmas, we derive 
that typable programs are deterministic.

\begin{lemma}\label{key-lemma2}
If $P\tact{\Gamma}{N} P_i$ for $i=1,2$ then 
$P_1 \twbis_{\uparrow(\Gamma)} P_2$.
\end{lemma}


\begin{theorem}[determinacy]\label{det-thm}
If 
$P \twact{\Gamma}{N} \cdot
  \twact{\Gamma'}{N} \cdots
  \twact{\Gamma'}{N} P_i, i=1,2, \Gamma'= \ucl \Gamma$
then $P_1\twbis_{\Gamma'} P_2$.
\end{theorem}

\section{Conclusion}
The main contribution of this work is the identification of 5 kinds of
usages in signal-based communication and of the  rules that allow
their {\em composition} while preserving determinacy. This goes well-beyond
previous analyses for {\sc Esterel}-like languages we are aware of that are
essentially `first-order' in the sense that signals are not treated as
first-class values.
Technically, we have shown that a typable process 
$P$ is {\em deterministic}.
This result builds on previous work by the authors \cite{Amadio06,AM07}
on a mathematical framework to reason about 
the equivalence of programs which is  comparable to the one available 
for the $\pi$-calculus.



\appendix

\section{Typing examples}\label{typing-examples}
We consider two examples that
are part of the folklore on synchronous
programming (see, {\em e.g.}, \cite{MandelPouzetPPDP05})
and a third one that suggests that a certain form
of single-assignment reference can be modelled in our framework.

\begin{example}[cell]\label{cell-ex}
We describe the behaviour of a generic cell that might be used in the
simulation of a dynamic system. Each cell relies on three parameters:
its state $q$,
its own activation signal $s$, and the list $\ell$ of
activation signals of its neighbours.
The cell performs the following operations in a cyclic fashion: (i) it
emits its current state along the activation
signals of its neighbours, (ii) it waits till the end of the current
instant (\s{pause}), and (iii) it collects the values emitted by its neighbours
and computes its new state.

{\footnotesize
\[
\begin{array}{lcll}
  \w{Cell}(q,s,\ell)&=& \w{Send}(q,s,\ell,\ell) \\
  \w{Send}(q,s,\ell,\ell')   &=&
  \matchv{\ell'}{\s{cons}(s',\ell'')} 
  {&(\emit{s'}{q} \mid \w{Send}(q,s,\ell,\ell''))}
  {\\&&&\s{pause}.\w{Cell}(next(q,!s),s,\ell)}
\end{array}
\]}

where $\w{next}$ is a function that computes the following state
of the cell according to its current state and the
state of its neighbours.
Assuming that the function $\w{next}$ is invariant under permutations
of the list of states, we would like to show that the evolution of
the simulation is deterministic. To express this invariance, 
a natural idea is to treat the `list' of distinct states as a `set',
{\em i.e.}, as a list {\em quotiented} by a relation that identifies
a list with any of its permutations.

We now turn to the typing.
Assume an inductive (non-affine) type
$\w{State}$ to represent the state of a cell and let
$\sigma= \w{Sig}_{u}(\w{State})$ where $u=(\infty,0,\infty)^\omega$
and $\sigma'= \w{List}_{\infty}(\sigma)$.
Then we can require:
$\w{Cell}:(\w{State},\sigma,\sigma')$ and 
$\w{Send}:(\w{State},\sigma,\sigma',\sigma')$.
Because, the usage of the signals under consideration is
$(\infty,0,\infty)^\omega$, the type of their dereferenciation is
$\w{Set}_{\infty}(\w{State})$ and therefore we must require
$\w{next}:(\w{State},\w{Set}_{\infty}(\w{State}))\arrow\w{State}$,
which means that the result of the function \w{next} must be invariant
under permutations of the list of (distinct) states.
\end{example}

\begin{example}[synchronous data flow]\label{data-flow-ex}
We provide an example of synchronous data-flow computation.
The network is described  by the program

{\footnotesize
\[
\nu s_2,s_3,s_4,s_5 (\ A(s_1,s_2,s_3,s_4) \mid B(s_2,s_3,s_5,s_6) \mid C(s_4,s_5)\ )
\]
\[
\mbox{where:}
\left\{
\begin{array}{ll}
A(s_1,s_2,s_3,s_4) &= s_1(x).(\emit{s_{2}}{f(x)} \mid s_3(y).
                      (\emit{s_{4}}{g(y)} \mid \s{pause}.A(s_1,s_2,s_3,s_4)),0),0 \\

B(s_2,s_3,s_5,s_6) &=s_2(x).(\emit{s_{3}}{i(x)} \mid s_5(y).(\emit{s_{6}}{l(y))} \mid \s{pause}.B(s_2,s_3,s_5,s_6)),0),0 \\

C(s_4,s_5)     &= s_4(x).(\emit{s_{5}}{h(x)}\mid \s{pause}.C(s_4,s_5)),0 

\end{array}\right.
\]}

Assuming that at each instant at most one value is emitted on the input signal
$s_1$, we would like to show that at each instant at most one value
will be emitted on every other signal.
This example suggests that we should introduce a notion of {\em affine
usage} in signals.

We now turn to the typing.
We assume an inductive type $D$ of {\em data} and let 
$\sigma = \w{Sig}_{u}(D)$,
$\sigma_I=\w{Sig}_{u_{I}}(D)$, and
$\sigma_O=\w{Sig}_{u_{O}}(D)$, where:
$u=(1,1,0)^\omega$,
$u_{I}=(0,1,0)^\omega$, and 
$u_{O}=(1,0,0)^\omega$. 
Then we can require:
$A: (\sigma_I,\sigma_O,\sigma_I,\sigma_O)$,
$B: (\sigma_I,\sigma_O,\sigma_I,\sigma_O)$, and
$C: (\sigma_I,\sigma_O)$.
The restricted signals $s_2,\ldots,s_5$ take the type $\sigma$ and the
overall system is well-typed with respect to the context
$s_1:\sigma_I,s_6:\sigma_O$. 
\end{example}

\begin{remark}[affinity vs. linearity]
With reference to the data flow example \ref{data-flow-ex}, one may notice that 
the type system guarantees determinacy by
making sure that at every instant {\em at most one} value is emitted on
every signal. One could consider a more refined type system that
guarantees that {\em exactly one} value is emitted on a signal at every
instant.\footnote{In this system the `else' branch of the input operator would become
useless} However, to obtain this system it is not enough to require
that all linear hypotheses in the context are used in the typing.
For instance, consider:
$\nu s,s':\sigma  (A(s,s') \mid A(s',s))$
where: $\sigma = \w{Sig}_{(1,1,0)^\omega}$,
$A:(\sigma,\sigma)$, and 
$A(s,s') = s().( \ol{s'} \mid \s{pause}.A(s,s') ),A(s,s')$.
This program could be {\em linearly} typed but it is stuck at every
instant. Following previous work (see, {\em e.g.}, \cite{K02}), 
one way to address this problem is to 
partition signals in a finite set of regions and to order them.
Then one designs typing rules that require 
that a reception on a signal belonging to a given region only
guards (prefixes) emissions on signals belonging to higher regions.
\end{remark}

\begin{example}[single-assign\-ment references]
We introduce a kind of {\em single-assign\-ment} references
that allow for a shared memory among different
threads while preserving determinacy. 
For simplicity, we look at
references on some basic inductive type $\kappa$. The three basic
operations are: (1)
$\s{newref} (s, e) \ P$ creates a reference $s$ whose
scope is $P$  and assigns it the value resulting from the evaluation of  $e$;
(2) $\s{read}(s,x).P$ reads the value $v$ contained in the 
reference $s$ and runs $[v/x]P$; and (3)
$\s{write}(s,e).P$ evaluates $e$ and writes its value in the
reference $s$. The written value will be available in the following
instant. Reading and writing are non-blocking operations, moreover a value
written at a given instant persists unless a following write
operation occurs.  To ensure determinacy, we have 
to guarantee that at any instant at most one value is written in a reference. 

We model this situation by associating with each reference $s$ a pair
of signals $(s,s')$. The first signal $s$ has a usage of kind 2 (one
write and arbitrarily many reads) while the signal $s'$ has a usage of
kind 5 (one write and one read during the instant). A reference $s$
containing the value $x$ is simulated by the following recursive
program:

{\footnotesize
\[
\begin{array}{ll}
\w{Ref}(s,s',x) &= 
\ol{s}x \mid s'(y).\s{pause}.\w{Ref}(s,s',y), \w{Ref}(s,s',x)
\end{array}
\]}

\noindent where the type of $\w{Ref}$ is 
$(\w{Sig}_{u}(\kappa), \w{Sig}_{u'}(\kappa),\kappa)$ with
$u= (1,\infty,\infty)^\omega$ and $u'=(0,1,0)^\omega$.
Thus on the signal $s$, $\w{Ref}$ emits the current
value of the reference while on the signal $s'$ it waits
for the value for the next instant. The usages we assign
to the signals $s$ and $s'$ guarantee that arbitrarily many
threads can read the reference but at most one can write it
at any given instant.
Formally, we can translate 
the three basic operations on references described above
as follows:

{\footnotesize
\[
\begin{array}{ll}

\la \s{newref} (s, e) \ P \ra 
&= 
\nu s, s'\ (\w{Ref}(s,s',e) \mid \la P \ra ), \\

\la \s{read}(s,x).P \ra 
&= 
s(x). \la P \ra,0, \\

\la \s{write}(s,e).P \ra 
&= 
\ol{s'}e \mid \la P \ra~.
 
\end{array}
\]}
\end{example}

\begin{example}[clocks]
We consider a kind of clock that 
still allows for a deterministic execution.\footnote{Note that in the usual
  semantics of timed automata, the fact that two processes may
  atomically {\em read and reset}  the same clock 
may produce race conditions.}
The value of a clock is a natural number which is
emitted on a signal, hence within an instant all threads 
can read the same clock value. 
At each instant, one or more threads may reset the clock value.
The effect of this reset is visible in the
following instant. To program a clock, we declare
the unit type and the type of natural numbers:

{\footnotesize
\[
\begin{array}{ll}
\w{Unit}_{\infty}() &= *  \\
\w{Nat}_{\infty}() &= Z \Alt S \ \w{of}\  \w{Nat}()
\end{array}
\]}

With each clock we associate a thread 
\w{Clock}  whose behaviour and type 
is defined as follows:

{\footnotesize
\[
\begin{array}{ll}
\w{Clock}(s,r,n) &= \ol{s}n \mid \s{pause}.\w{Clock'}(s,r,!r,n) \\
\w{Clock}        &:
(\w{Sig}_{u}(\w{Nat}),\w{Sig}_{u'}(\w{Unit}),\w{Nat}),\quad
                 u=(1,\infty,\infty)^\omega, \quad u'=(\infty,0,1)^\omega \\
\w{Clock'}(s,r,\ell,n)    
&=\matchv{\ell}{\s{nil}}{\w{Clock}(s,r,S(n))}{\w{Clock}(s,r,Z)} \\
\w{Clock'}       &: (\w{Sig}_{u}(\w{Nat}),\w{Sig}_{u'}(\w{Unit}),
\w{Set}_{1}(\w{Unit}),
\w{Nat})
\end{array}
\]}

Note that the typing guarantees that the thread $\w{Clock}$
is the only one that can emit the clock signal $s$ and read
the reset signal $r$. On the other hand, another thread using the
clock may {\em read} the clock value on the signal $s$ and 
may reset it in the following instant by emitting on 
the reset signal $r$.
\end{example}

\section{Proofs}

\subsection{Proof of lemma \ref{weak-lemma}}
By induction on the typing rules. One uses several times
the fact that $\oplus$ is associative and commutative both
on types and contexts and the fact that the rules are
formulated so that the conclusion still holds when
the usages in the context $\Gamma$ are increased 
(see, {\em e.g.}, the rule $(\w{var})$).

\subsection{Proof of lemma \ref{sub-lemma}}
The following lemma collects some preliminary remarks.

\begin{lemma}\label{aux-sub-lemma}
\Defitemf{(1)} 
If  $\Gamma \Gives U:T$, $\Gamma'\Gives v:\rho$, 
$(\Gamma \oplus \Gamma')\dcl$, and $x\notin\w{dom}(\Gamma)$
then \\ \noindent $(\Gamma \oplus \Gamma') \Gives [v/x]U:T$.

\Defitem{(2)} 
If $\Gamma \Gives v:\kappa$ then there is a neutral context 
$\Gamma'$ such that $\Gamma'\Gives v:\kappa$ and 
$\Gamma = \Gamma'\oplus \Gamma''$.

\Defitem{(3)} If $\Gamma \Gives v:\rho$ and $\rho = \rho_1\oplus \cdots \oplus \rho_n$ then there exist $\Gamma_1,\ldots,\Gamma_n$ such that 
$\Gamma_1\oplus \cdots \oplus \Gamma_n=\Gamma$ and $\Gamma_i\Gives v:\rho_i$
for $i=1,\ldots,n$.
\end{lemma}
{\sc Proof} 
\Proofitemf{(1)}
If $x\in \w{FV}(U)$ then the only possibility is that $x\in \w{FV}(e)$ where
$\ul{\ol{s}{e}}$ is a sub-term of $U$. But then 
one can type $\ul{\ol{s}{[v/x]e}}$ exactly as one types $\ul{\ol{s}{e}}$.
So $\Gamma \Gives [v/x]U:T$ and we conclude by weakening.

\Proofitem{(2)} We proceed by induction on $v$.
For the inductive step, we use the fact that if 
$\s{c}(v_1,\ldots,v_n)$ has a neutral type then
the $v_i$ must have a neutral type too.

\Proofitem{(3)} If the type $\rho$ is neutral then 
$\rho=\rho_1=\cdots = \rho_n$. By (2), we can find a neutral
context $\Gamma'$ such $\Gamma'\Gives v:\rho$ and $\Gamma'\oplus \Gamma''=\Gamma$. Then it suffices to take $\Gamma_1= \Gamma'\oplus \Gamma''$ and 
$\Gamma_i= \Gamma'$ for $i=2,\ldots,n$.
If the type $\rho$ is affine and either an inductive type or a set type then
we must have $n=1$ and the assertion follows immediately.
Finally, if the type $\rho$ is affine and a signal type then the usages
of the signal in the types $\rho_1,\ldots,\rho_n$ allow to construct directly
the contexts $\Gamma_1,\ldots,\Gamma_n$. \qed  

Next, to prove the substitution lemma we proceed by
induction on the typing of $U$.

\Proofitem{(\w{var})}
Suppose $\Gamma,y:\w{Op}_{u}(\sigma) \Gives y:\w{Op}_{u'}(\sigma)$ with $u\geq
u'$.

\Defitem{\bullet}
If $\Gamma = \Gamma'',x:\rho$ and $x\neq y$ then
$((\Gamma'',y:\w{Op}_{u}(\sigma))\oplus \Gamma')(y) = \w{Op}_{u''}(\sigma)$ 
with $u''\geq u$. Hence, by $(\w{var})$,
$(\Gamma'',y:\w{Op}_{u})\oplus \Gamma' \Gives y:\w{Op}_{u'}$.

\Defitem{\bullet}
If $x=y$ then $[v/x]y=v$.
If $\w{Op}$ is not $\w{Sig}$ then $u=u'$. 
By hypothesis,
$\Gamma '\Gives v: \w{Op}_{u}(\sigma)$ and by weakening
$\Gamma''\oplus \Gamma' \Gives v: \w{Op}_{u}(\sigma)$.
On the other hand, if $\w{Op}$ is $\w{Sig}$ then, 
by $(\w{var})$,  $(\Gamma''\oplus \Gamma')\Gives v:\w{Op}_{u}(\sigma)$.

\Proofitem{(k)} 
If $k$ is a constant then apply weakening. Otherwise,
suppose $\Gamma,x:\rho = \Gamma_0\oplus \Gamma_1 \oplus \cdots \oplus
\Gamma_n$ with $\Gamma_i \Gives e_i:\sigma_i$, $i=1,\ldots,n$.
Let $I=\set{i\in\set{1,\ldots,n} \mid x\in \w{dom}(\Gamma_i)}$.
If $i\in I$ then assume $\Gamma_i= \Gamma''_i,x:\rho_i$. 
We have $\rho = \oplus_{i\in I} \rho_i$. 
By lemma \ref{aux-sub-lemma}(3), we can find $\Gamma'_i$ such that
$\Gamma'_i \Gives v:\rho_i$ for $i\in I$ and $\Gamma'= \oplus_{i\in I}
\Gamma'_i$.  If $i\notin I$ then $\Gamma_i \Gives [v/x]e_i:\sigma_i$,
(cf.  lemma \ref{aux-sub-lemma}(1)), and if $i\in I$ then 
$(\Gamma_i\oplus \Gamma'_i) \Gives [v/x]e_i:\sigma_i$, 
by inductive hypothesis. 

\medskip\noindent
This kind of argument is repeated several times for the
remaining rules. As already pointed out in the proof 
of the weakening lemma \ref{weak-lemma},
another important point is that the rules
are built so that adding extra capabilities to the hypotheses in 
the context does not affect the conclusion. We just look in some detail 
at the rule $[\w{var}_{\w{sig}}]$ in the case where
$\Gamma,s:\w{Sig}_{xy^{\omega}}(\sigma) \Gives [s]:\w{Sig}_{u}(\sigma)$,
$y^\omega \geq u$, $\Gamma'\Gives s':\w{Sig}_{xy^\omega}(\sigma)$ and 
$(\Gamma \oplus \Gamma')\dcl$. 
Then $\Gamma'(s) = s':\w{Sig}_{u'}(\sigma)$ with $u'\geq xy^\omega$.
Hence $\uparrow(u') \geq y^\omega \geq u$. \qed


\subsection{Proof of lemma \ref{exp-eval-lemma}}
By induction on the evaluation $e\eval v$.
If $e$ is a signal $s$ or a constant $\s{c}$ then
$e=v$ and the conclusion is immediate. So suppose:
$e=k(e_1,\ldots,e_n)$, 
$k:(\sigma_1,\ldots,\sigma_n)\arrow \sigma$,
$\Gamma= \Gamma_0\oplus \Gamma_1\oplus \cdots \oplus \Gamma_n$, 
$\Gamma_i \Gives e_i:\sigma_i$, and 
$e_i\eval v_i$, for $i=1,\ldots,n$.
By inductive hypothesis, $\Gamma_i \Gives v_i:\sigma_i$, for 
$i=1,\ldots,n$. If $k$ is a constructor $\s{c}$ then
$v=\s{c}(v_1,\ldots,v_n)$ and $\Gamma \Gives v:\sigma$ by
the rule $(k)$. If $k$ is a function $f$ then 
again by the rule $(k)$, $\Gamma \Gives f(v_1,\ldots,v_n):\sigma$ and,
by hypothesis on $f$, we have that $f(v_1,\ldots,v_n)\eval v$ and
$\Gamma \Gives v:\sigma$. \qed

\subsection{Proof of lemma \ref{sub-eoi}}

\Proofitem{(1)} The effect of $V(A(\vc{r}))$ is to replace each
of occurrence of $!s$ in $\vc{r}$ with $V(s)$. 
First notice that if $!s$ occurs in $\vc{r}$ then its usage cannot
be of kind $5$. Moreover, if it is of kind $1$ or $2$ then we can
have several occurrences of $!s$ in $\vc{r}$ and the type
of the values emitted on the signal must be non-affine. 
Notice that to type a non-affine value, we just need a non-affine context
and since non-affine types are (exactly the) neutral types, we can
use this context as many times as needed.
On the other hand, if the signal is of kind $3$ or $4$ then
the values emitted on the signal can be affine but
there can be no more than one occurrence of $!s$ in $\vc{r}$.

Following these preliminary considerations, we proceed by 
case analysis on the rules $[!_{\w{Set}}]$ and $[!_{\w{List}}]$.
In each case, one has a judgement of the shape:
\[
\Gamma,s: \w{Sig}_{u}(\sigma) \Gives [!s]: \w{Op}_{x}(\sigma)
\]
knowing that $\Gamma'\Gives V(s)=[v_1;\ldots;v_n]:\w{Op}_{x}(\sigma)$,

\Proofitem{(2)}
By definition, $V(A(r_1,\ldots,r_n)) = A(V(r_1),\ldots,V(r_n))$.
Suppose $A:(\sigma_1,\ldots,\sigma_n)$. We know that
$v_i\sim_{\sigma_{i}} u_i$ entails that $A(v_1,\ldots,v_n)\wbis 
A(u_1,\ldots,u_n)$. Hence, it is enough to show that
that $V(r_i) \sim_{\sigma_{i}} V'(r_i)$ for $i=1,\ldots,n$. 
We proceed by induction on the structure of $r$.
If $r$ is a signal or a constant then by definition
$r\sim_{\sigma_{i}} r$.
If $r$ is of the shape $!s$ then we analyse the kind of usage of $s$.
If it is of kind $2$ or $4$ then $V(s)=V'(s)$ (there is at most
one value in the lists).
If it is of kind $1$ or $3$ then $V(s)$ and $V'(s)$ are equal
up to permutation, and we rely on the definition of $\sim$
on set types.
Finally, if $r=k(\vc{r})$ we apply the inductive hypothesis
plus the definition of $\sim$ on constructors if $k$ is a constructor
and the hypothesis on the functions if $k$ is a function.

\subsection{Residual context on auxiliary actions}\label{residual-aux-sec}
We specify the notion of {\em residual context} 
on auxiliary actions. The definition for the actions $s?v$ is
similar to the one for the actions $sv$. On the other hand, for
the actions $(E,V)$,  we have to analyse how a program exports and 
imports usages at the end of the instant. For instance, consider
$P=\emit{s_{1}}{t_{1}} \mid \emit{s_{2}}{t_{2}} \mid A(!s_1)$,
and suppose $P \tact{\Gamma}{(E,V)} A(V(s_{1}))$ where:
\[
\begin{array}{llllll}
E &= &[\set{t_{1}}/s_{1},\set{t_{2}}/s_{2}] \qquad
&V &= &[[t_1;t_3]/s_{1},[t_4;t_2]/s_{2}] ~.
\end{array}
\]
The function $E$ represents what $P$ emits, the function $V$
represents what $P$ assumes to be emitted, moreover looking at the
context $\Gamma$, we may determine what the process $P$ may receive
at the end of the instant (note that $P$ may receive what it
emits and that a value with an affine typing can be received
at most once). In computing the residual context, we have 
to subtract what is exported to the environment while 
adding what is imported from it. 
Going back to our example, clearly the context $\Gamma$ must specify
that $P$ may receive on $s_1$ at the end of the instant. Suppose
moreover that it specifies that $P$ may not receive on $s_2$. Then in
computing the residual context, we have to subtract the usage for
$t_2$ which is exported to the environment while adding 
the usage for $t_3$ which is received from it.
Following these considerations, we define:

{\small
\[
\begin{array}{llll}

\Delta(E,\Gamma) 
&=
&\oplus \set{\Delta(v,\lambda) \mid
\Gamma(s)=\w{Sig}_{u}(\lambda), 
v\in E(s),
u(0)_3\neq 1}
&(\mbox{export}) \\ \\

\Delta(V,\Gamma) 
&= 
&\oplus \set{\Delta(v,\sigma) \mid 
\Gamma(s)=\w{Sig}_{u}(\sigma),
v \in V(s), 
u(0)_3\neq 0}
&(\mbox{import})

\end{array}
\]}

Note that in the `exported context' $\Delta(E,\Gamma)$ we only care
about usages of values of affine type, as otherwise
$\Delta(v,\kappa)$ is neutral.  On the other hand, in the `imported
context' we look at all the values regardless of their type.  Indeed,
$v$ might have a neutral type but contain a fresh signal name and then
we need to import a neutral context to type it.  Also note that in the
following definition \ref{res-aux-def}, we actually focus only on the
values that are {\em not} emitted (in $E$).

\begin{definition}[residual context on auxiliary actions]\label{res-aux-def}
Given a context $\Gamma$ and an auxiliary action $\w{aux}$ the
residual context $\Gamma(\w{aux})$ is defined as follows where
$u_{5}$ is as in definition \ref{res-cxt-def}:

{\small
  \[
  \Gamma(\w{aux}) = \left\{
    \begin{array}{ll}
     
 (\Gamma \ominus\set{s:Sig_{u_{5}}(\sigma')})\oplus\Delta(v,\sigma')
      & \mbox{if }\Gamma(s) = Sig_u(\sigma'), \w{aux}=s?v, \mbox{ and }(1)\\

      (\ucl\Gamma  \ominus \Delta(E, \Gamma))  \oplus   \Delta(V',  \Gamma)   

& \mbox{if }\w{aux}=(E,V)\mbox{ and }V\minus E=V'\\
    \end{array}
  \right.
  \]}
\end{definition}

\subsection{Proof of theorem \ref{subj-red-thm}}
We proceed by induction on the proof of the transition and 
by case analysis on the action $\w{act}$ which is performed.

\Proofitem{(sv)} There is just 1 rule to consider: $(\w{in})$.
Suppose $\Gamma(s) = \w{Sig}_u(\sigma')$. 
The definition of the residual context provides an 
additional context $\Delta(v,\sigma')\oplus\set{s:\w{Sig}_{u_{out}}(\sigma')}$
which is just what is needed to type $\emit{s}{v}$.

\Proofitem{(s?v)} There are 3 rules to consider: $(\w{in}_{\w{aux}})$,
$(\w{comp})$, and $(\nu)$. We just look at the first one.
Suppose $(\Gamma_1\oplus \Gamma_2) \Gives s(x).P,K$, 
$\Gamma_1 \Gives s:\w{Sig}_u(\sigma')$, $u(0)_2\neq 0$, 
$\Gamma_2,x:\sigma'\Gives P$, and $\Gamma_1\oplus \Gamma_2\Gives [K]$.
Note that necessarily $u\geq u_{\w{in}}$.
By construction, $\Delta(v,\sigma') \Gives v:\sigma'$. 
By the substitution lemma \ref{sub-lemma}, 
$\Gamma_2 \oplus \Delta(v,\sigma') \Gives [v/x]P$
and then it is enough to apply weakening to get the residual context.

\Proofitem{(\nu \vc{t:\sigma} \ \emit{s}{v})}
There are 5 rules to consider: $(\w{out})$, with a special treatment for
kind 5, $(\ul{\w{out}})$, $(\nu_{\w{ex}})$, $(\w{comp})$, and $(\nu)$.

\Proofitem{(\tau)}
There are 8 rules to consider: 
$(\w{synch})$, $(\w{rec})$, $(=_{i}^{\w{sig}})$, 
$(=_{i}^{\w{ind}})$, $(\w{comp})$, and $(\nu)$ for
$i=1,2$ We just look at the first two.

\Proofitem{(\w{synch})}
Suppose: 
$P_1 \act{\nu \vc{t:\rho} \emit{s}{v}} P'_1$, 
$P_2 \act{s?v} P'_2$, $\Gamma_i\Gives P_i$, for $i=1,2$,
and $(\Gamma_1\oplus \Gamma_2)(s)= \w{Sig}_{u}(\sigma')$.
By inductive hypothesis, we have:
\[
\begin{array}{l}
(\Gamma_1,\vc{t:\rho}) \ominus \Delta(v,\sigma')\oplus
\set{s:\w{Sig}_{u_{5}}(\sigma')} \Gives P'_1\quad \mbox{and} \\

(\Gamma_2 \oplus \Delta(v,\sigma') \ominus 
\set{s:\w{Sig}_{u_{5}}(\sigma')} \Gives P'_2 
\end{array}
\]
Recall that here $u$ may be of kind 2 or 5 and that in the first
case $u_5$ is neutral. In both cases, we get
$(\Gamma_1 \oplus \Gamma_2),\vc{t:\rho} \Gives (P'_1\mid P'_2)$,
and we conclude applying the typing rule $(\nu)$.

\Proofitem{(\w{rec})}
Suppose $A:(\sigma_1,\ldots,\sigma_n)$, $\Gamma_i\Gives e_i:\sigma_i$,
$e_i\eval v_i$, for $i=1,\ldots,n$. 
By lemma \ref{exp-eval-lemma}, $\Gamma_i\Gives v_i:\sigma_i$.
By hypothesis, we know that if $A(x_1,\ldots,x_n)=P$ then
$x_1:\sigma_1,\ldots,x_n:\sigma_n \Gives P$. Thus, by iterating
the substitution lemma \ref{sub-lemma}, we get, as required,
$\Gamma_1\oplus \cdots \oplus \Gamma_n \Gives [v_1/x_1,\ldots,v_n/x_n]P$.

\Proofitem{(E,V)}
There are $5$ rules to consider:
$(0)$, $(\w{reset})$, $(\ul{\w{reset}})$, $(\w{cont})$, and
$(\w{par})$. We focus on the last two.

\Proofitem{(\w{cont})}
Suppose $s(x).P,K \act{(\emptyset,V)} V(K)$ and
$\Gamma \Gives s(x).P,K$. Then $\Gamma \Gives [K]$.
We rely on lemma \ref{sub-eoi}(1). 
We build the context $\Gamma'$ in the lemma by 
taking $\Gamma'=\Delta(V,\Gamma)$ which is uniform added
to a context $\Gamma''$ which just provides the usages to emit
in the first instant the values in $V$ on the signals in $\w{dom}(V)$.

\Proofitem{(\w{par})}
Suppose: $\Gamma=(\Gamma_1\oplus \Gamma_2)$, $\Gamma \Gives (P_1\mid P_2)$,
$(P_1\mid P_2) \act{(E_1\union E_2),V} (P'_1\mid P'_2)$, $\Gamma_i
\Gives P_i$, $P_i \act{(E_i,V)} P'_i$, for $i=1,2$.
Following the definition of residual context, define for $i=1,2$:
{\small
\[
\begin{array}{ll}

\w{Exp}_i = \Delta(E_i,\Gamma_i)
&\w{Exp}_{1,2} = \Delta(E_1\union E_2,\Gamma_1\oplus \Gamma_2) \\

\w{Imp}_i = \Delta(V\minus E_i,\Gamma_i)
&\w{Imp}_{1,2} = \Delta(V\minus(E_1\union E_2),\Gamma_1\oplus \Gamma_2) \\

\Gamma'_i = \uparrow \Gamma_i \ominus \w{Exp}_i \oplus \w{Imp}_i  

&\Gamma' = \uparrow(\Gamma_1\oplus \Gamma_2) \ominus \w{Exp}_{1,2} \oplus
\w{Imp}_{1,2} 

\end{array}
\]}
We want to show $\Gamma'= \Gamma'_1\oplus \Gamma'_2$.  
We proceed, by analysing the contribution of each value
$v\in V(s)$ such that $\Gamma(s)=\w{Sig}_{u}(\sigma)$
to the computation of $\w{Imp}_i$, $\w{Imp}_{1,2}$, $\w{Exp}_{i}$, and
$\w{Exp}_{1,2}$. We use the notation, {\em e.g.}, 
$\w{Imp}_{1}(v)$ to denote the contribution of the value $v$
to the computation of the context $\w{Imp}_{1}$.

\Defitem{\bullet}
If $\sigma$ is non-affine then, for $i=1,2$, 
$\w{Imp}_{i}$, and $\w{Imp}_{1,2}$ are neutral contexts
while $\w{Exp}_{i}$ and $\w{Exp}_{1,2}$ are empty contexts.
Up to symmetries, 
$v$ can be received either by (i) $\Gamma_i$, $i=1,2$ or 
(ii) $\Gamma_1$ and $\Gamma_2$ and emitted either
by (i) $E_1\inter E_2$, or (ii) $E_1\minus E_2$,
or (iii) $E_2\minus E_1$, or by (iv) the environment.
One proceeds by case analysis (8 situations).

\Defitem{\bullet}
If $\sigma$ is affine then the usage $u$ must be of kind $3$ or $4$
and at the end of the instant the signal $s$ may be read, exclusively,
either by (i) $\Gamma_i$, $i=1,2$ or by (ii) the environment.  On the other
hand, $v$ may be emitted either by (i) $(E_1\inter E_2)$, or by (ii)
$(E_1\minus E_2)$, or by (iii) $(E_2\minus E_1)$ or by (iv)
$(V\minus (E_1\union E_2))$.  If $v\in (E_1\inter E_2)(s)$ then 
$\Delta(v,\sigma)$ must be neutral for otherwise the addition
is not defined. One then proceeds by case analysis 
(8 situations). Note that if the environment receives $v$ then the import
contexts $\w{Imp}_{i},\w{Imp}_{1,2}$ are empty while if
$\Gamma_i$ receives $v$ then $\w{Exp}_i$ is empty.

\Proofitem{(N)}
There is just $1$ rule to consider: $(\w{next})$. 
Suppose $\Gamma \Gives P$ and $P\nor \nu \vc{\vc{s}:\rho} \ P''$. 
Clearly, a typing of, say, $(\nu s:\rho\  Q_1) \mid Q_2$ can be transformed
into a typing of $\nu s:\rho \ (Q_1\mid Q_2)$. Thus
$\Gamma \Gives \nu \vc{\vc{s}:\rho} \ P''$ and
$\Gamma,\vc{s}:\rho \Gives P''$. 
By definition of the rule $(\w{next})$, $P''\act{(E,V)} P'$ with $V \real E$.
By inductive hypothesis and weakening, 
$\uparrow (\Gamma,\vc{s:\rho}) \Gives P'$.
Thus $\uparrow(\Gamma) \Gives \nu \vc{s:\uparrow\rho}\ \Gives P'$. \qed

\subsection{Proof of proposition \ref{free-typed-bis}}\label{proof-free-typed-bis}
We show that the following indexed relation is a typed bisimulation:
\[
P \trl{R}{\Gamma} Q \quad \mbox{if}
\quad P,Q\in \w{Pr}(\Gamma) \mbox{ and } P\wbis Q~.
\]
Suppose $P\trl{R}{\Gamma} Q$, $P\tact{\Gamma}{\alpha} Q$, and 
$\w{bn}(\alpha) \inter \w{fn}(Q)=\emptyset$.  Then:

{\small\[
\begin{array}{ll}

P\act{\alpha} P'               &\mbox{(by definition of typed transition)} \\
\Gamma(\alpha) \Gives P'       &\mbox{(by subject reduction)} \\
Q\wact{\alpha} Q', P'\wbis Q'  &\mbox{(by untyped bisimulation)} \\
\Gamma(\alpha) \Gives Q'      &\mbox{(by subject reduction)}

\end{array}
\]}

Hence we can conclude that $P'\trl{R}{\Gamma(\alpha)} Q'$. \qed

\subsection{Proof of lemma \ref{key-lemma1}}
\Proofitemf{(1)} 
An inspection of the labelled transition system in table
\ref{Lts} reveals that two $\tau$ reductions may superpose
only if they are produced by two synchronisations on the same
signal name, say $s$. In this case, $s$ must have a usage
of kind 2 or 5. In a usage of kind 2, the typing guarantees
that there is at most one value emitted on $s$ so that 
we are roughly in the following situation:
\[
P=C[s(x).P_1,Q_1 \mid s(x).P_2,Q_2 \mid \emit{s}{e}]
\] 
Because a signal emission persists within an instant, 
it is possible to close the diagram in one step.
On the other hand, in a usage of kind 5 there can be at most
one receiver and therefore no superposition may arise.

\Proofitem{(2)} We show that $\tacteq{\Gamma}{\tau}$ is 
a typed bisimulation. If $P=Q$ nothing needs to be proved.
So suppose $P \tact{\Gamma}{\tau} Q$. Clearly, $P$ can weakly
simulate all actions $Q$ may perform just by performing initially
an extra $\tau$ step. So suppose $P \tact{\Gamma}{\alpha} P'$.
Note that $\alpha \neq N$ since $P$ may perform a $\tau$ action.

\Defitem{\alpha=\tau} 
In this case, we apply (1) noticing that $\tacteq{\Gamma}{\tau} \subseteq
\twact{\Gamma}{\tau}$.

\Defitem{\alpha = sv}
In this case, $P'= (P \mid \emit{s}{v})$ and we can close the diagram
by performing $Q\act{sv} (Q\mid \emit{s}{v})$.

\Defitem{\alpha= \nu \vc{t} \emit{s}{v}}
Again, because a value emitted on a signal persists, it is equivalent
to use it in an internal synchronisation and then again to extrude
the value to the environment or the other way around. \qed

\subsection{Proof of lemma \ref{key-lemma2}}
By subject reduction we know that $\ucl(\Gamma) \Gives P_i$.
If we can show that $P_1\wbis P_2$ then by 
proposition \ref{free-typed-bis} we can conclude.
According to the rule $(\w{next})$ of the labelled transition system,
we must have for $i=1,2$:
\[
P\nor \nu \vc{s}_{i}\  P',
\quad \vc{s}_{1} \mbox{ permutation of }\vc{s}_{2},  
\qquad
P' \act{E,V_i} P''_{i}, \quad V_i\real E, 
\qquad
P_i = \nu \vc{s}_{i} P''_{i}~.
\]
Then lemma \ref{sub-eoi}(2) and fact \ref{bis-cong-fact} guarantee that
$P''_{1} \wbis P''_{2}$ and $P_{1} \wbis P_2$. \qed

\subsection{Proof of theorem \ref{det-thm}}
The proof is a direct diagram chasing relying on lemma \ref{key-lemma1}(2),
\ref{key-lemma2}, and the definition of typed bisimulation. \qed

\end{document}